%% file: ms.tex
\documentclass{emulateapj}
\usepackage{graphicx,epsfig,times}
\newcommand {\kms}{km s$^{-1}$}

\begin{document}
  \title{Optical Selection of Faint AGN in the COSMOS field}
  \shorttitle{Optical Selection of Faint AGN in COSMOS}
  \shortauthors{Casey et al.}
  \author{C. M. Casey\altaffilmark{1,2}, C. D. Impey\altaffilmark{1}, 
    J. R. Trump\altaffilmark{1}, J. Gabor\altaffilmark{1}, R.G. Abraham\altaffilmark{3},
    P. Capak\altaffilmark{4}, N.Z. Scoville\altaffilmark{4}, M. Brusa\altaffilmark{5},
    E. Schinnerer\altaffilmark{6}}
  \altaffiltext{1}{Steward Observatory, University of Arizona, 933 N. Cherry St, Tucson, AZ 85721, U.S.A.}
  \altaffiltext{2}{Institute of Astronomy, University of Cambridge, Madingley Rd, Cambridge, CB3 0HA, U.K.}
  \altaffiltext{3}{Department of Astronomy and Astrophysics, University of Toronto, 50 St. George St, Toronto, 
    Ontario, M5S 3H4, Canada}
  \altaffiltext{4}{Astronomy Department, California Institute of Technology, 1200 East California Blvd, Pasadena, CA, 91125, U.S.A.}
  \altaffiltext{5}{Max-Planck-Institut f\"{u}r extraterrestrische Physik, Postfach 1312, 85741 Garching, Germany}
  \altaffiltext{6}{Max-Planck-Institut f\"{u}r Astronomie, K\"{o}nigstuhl, D-69117 Heidelberg, Germany}

  \begin{abstract}
    We outline a strategy to select faint ($i_{AB} < 24.5$) type 1 AGN candidates down to the 
    Seyfert/QSO boundary for spectroscopic targeting in the COSMOS field \citep{scoville07a}.  
    Our selection process picks candidates by their nonstellar colors in $uBVRizK$ broadband 
    photometry from the Subaru and CFH Telescopes and morphological properties extracted from 
    HST ACS $i$ band data.  Although the COSMOS field has been used extensively to survey the 
    faint galaxy population out to $z \sim 6$, AGN optical color selection has not been applied 
    to so faint a level in such a large continuous part of the sky.  Hot stars are known to be 
    the dominant contaminant for bright AGN candidate selection at $z < 2$, but we anticipate 
    the highest color contamination rate at all redshifts to be from faint starburst and compact 
    galaxies.  Morphological selection via the Gini Coefficient separates most potential AGN from 
    these faint blue galaxies.  Recent models of the quasar luminosity function (QLF) from 
    \citet{hopkins07a} are used to estimate quasar surface densities, and a recent study of 
    stellar populations in the COSMOS field \citep{robin07a} is applied to infer stellar surface 
    densities and contamination.  We use 292 spectroscopically confirmed type 1 broad line AGN 
    and quasar templates to predict AGN colors as a function of redshift, and then contrast 
    those predictions with the colors of known contaminating populations.  Since the number of
    galaxy contaminants cannot be reliably identified with respect to stellar and predicted 
    QLF numbers, the completeness and efficiency of the selection cannot be calculated before 
    gathering confirming spectroscopic observations.  Instead we offer an upper limit estimate 
    to selection efficiency (about 50$\%$ for low-z and 20-40$\%$ for int-z and high-z) as well 
    as the completeness and efficiency with respect to an X-Ray point source population (from 
    the COSMOS AGN Survey), in the range 20$\%$ to 50$\%$.  The motivation of this study and 
    subsequent spectroscopic follow up is to populate and refine the faint end of the QLF, at 
    both low and high redshifts, where the population of type 1 AGN is presently not well known.  
    The anticipated AGN observations will add to the $\sim$300 already known AGN in the COSMOS 
    field, making COSMOS a densely packed field of quasars to be used to understand supermassive black 
    holes and probe the structure of the intergalactic medium in the intervening volume.
  \end{abstract}

  
  \section{INTRODUCTION}
  
  Optical colors provide a well-developed, reliable astronomical selection technique for stellar and 
  galaxy populations.  The method was first applied to AGN in the 1960's, based on the inference that 
  quasars often have a larger ultraviolet excess than the hottest stars \citep{sandage65a}.  Subsequent 
  large-scale surveys have taken up the search for quasars \citep[e.g.][]{schmidt83a,foltz87a,croom01a,schneider07a}, 
  causing the known population to grow dramatically.  The ongoing search to find new quasars is highly motivated
  by their use in probing the intergalactic medium (IGM) and understanding the nature of supermassive black holes.
  To efficiently target and identify new quasars, optical selection techniques have proven to be highly
  efficient, in some cases mitigating the need for confirming slit spectroscopy \citep{richards02a,richards04a}.  
  \citet{richards02a} used multi-color imaging from the Sloan Digital Sky Survey (SDSS) to select AGN 
  and quasars down to magnitudes $i_{AB} < 21$.  In this paper we apply optical selection to the COSMOS 
  field \citep{scoville07a}, probing the AGN population to much fainter magnitudes ($i_{AB} < 24.5$) than 
  any previous large-area survey, and we reveal challenges unique to the fainter AGN population and its 
  contaminants.  To properly account for contamination of the AGN candidate pool, we characterize the 
  stellar populations that are dominant at $i_{AB} < 21$ and the galaxy population that are more prevalant 
  at fainter magnitudes.

  Targeting the AGN population to such a faint level is key to understanding bulk properties of AGN and 
  constraining the faint end of the quasar luminosity function (QLF) at high redshift, which is highly 
  unknown and can vary in up to two orders of magnitude at $i > 23$ \citep*[e.g. pure luminosity evolution 
  vs. luminosity dependent density evolution most recently presented in][]{hopkins07a}.  With a more complete 
  QLF, astronomers can analyze the nature of low luminosity quasars further $-$ answering important questions 
  about their host galaxies and environments.  Such objects are also useful for interpreting the low mass end 
  of the black hole M-$\sigma$ relation, and for probing the IGM.  A particular goal of observing faint AGN 
  is to measure the growth rate of lower mass black holes and/or AGN that accrete with lower efficiency.  
  This faint survey brings quasar selection into a new regime of luminosity, placing new observational 
  bounds on theoretical ideas about the nature and evolution of quasars.
  
  \citet{richards04a} used two 3D multi-color spaces to select QSO candidates in SDSS: $ubri$ $(u-b,b-r,r-i)$ 
  for lower redshift candidates and $briz$ $(b-r,r-i,i-z)$ for candidates with $z > 2.5$.
  The AGN population with $2.5 < z < 3.0$ is well known to exhibit colors similar to A stars and thus 
  is extremely difficult to isolate via optical means \citep{richards02a,fan99a,richards01a}, justifying 
  a split of the selection algorithm into high and low redshift components.  Following the SDSS group, we use 
  a $u-B$ baseline for $z < 2.5$ and a combination of $(B-V)$ and $(V-i)$ colors to target AGN with $z > 3.0$.  
  The intermediate redshift range ($2.5 < z < 3.0$) is also targeted and follows similar selection criteria as 
  the high redshift selection, but we expect a much lower object yield in this range due to heavy contamination 
  from faint blue stars.  Unlike the SDSS group, we do not anticipate recovering the AGN population with equal 
  efficiencies across all redshifts. Our goal is to push AGN selection to fainter magnitudes while maintaining 
  reasonable efficiency ($>20\%$) and completeness ($>30\%$).  Unique to our survey is the use of morphological 
  information (via the ACS images) to separate the marginally resolved AGN galaxies and unresolved stars (only 
  prominent in number at the brighter magnitudes) from more clearly resolved galaxies.

  Since the goal of this study is to realistically constrain the faint AGN population, we hope to target a 
  significant portion of our AGN candidates during future spectroscopic ovservations, anticipating anywhere 
  from 30-50 AGN yeild per night by observing $\sim$100 candidates, as well as gaining important information 
  from the spectroscopic details of contaminating galaxies.  Already, $\sim$160 candidates, chosen by the 
  methodology of this paper, have been observed at Magellan IMACS and LDSS3 as of May 2007 and more observations 
  are planned.  By building up significant statistics on low-luminosity AGN in COSMOS, such a large swath of the 
  sky, we are in a unique positions to improve what is known about the AGN population with meaningful statistics 
  at the limit of current observations.  

  This paper thoroughly discusses the development of a reliable optical AGN selection algorithm, current knowledge
  of the QLF, and estimates of our algorithm's efficiency and completeness; observations and further development
  of the QLF will be discussed in a follow-up paper.  The catalogs and data used in the development of a selection 
  algorithm are discussed in \S \ref{catalog_s}.  The colors and nature of the contaminating populations are discussed 
  in \S \ref{colorcontam_s}, while our method of morphological selection is given in \S \ref{morphsel_s}.  The specifics 
  of the AGN selection are detailed in \S \ref{algorithm_s}.   In \S \ref{qlf_s} we discuss the current picture of the QLF,
  predict number counts of contaminating populations, and discuss estimates to the efficiency of our methodology.  
  We use a standard cosmology with $\Omega_{M} = 0.3$, $\Omega_{\Lambda} = 0.7$, and $H_{0} = 70$ \kms Mpc$^{-1}$ 
  \citep[e.g.][]{spergel03a} and luminosity distances computed according to \citet{hogg99a}.

  \section{CATALOGS AND TRAINING DATA}\label{catalog_s}

\begin{figure}
    \centering
    \includegraphics[width=0.99\columnwidth]{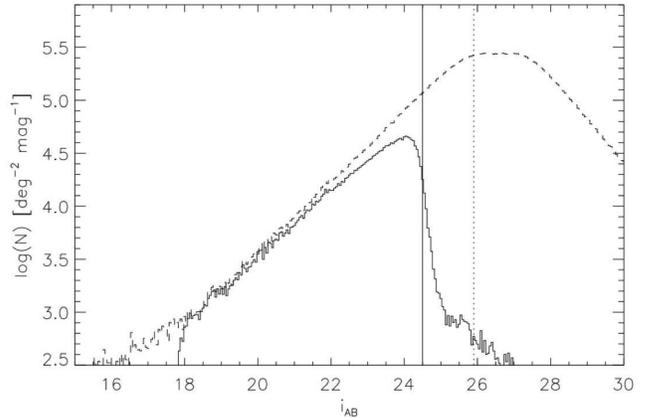}
    \caption{
      Differential distribution of $i_{AB}$ apparent magnitude (in a bin size of 0.05 mag) for all objects in the COSMOS 
      photometric (dashed) and morphological (solid) catalogs.  The larger COSMOS Photometric Catalog (CPC) contains over 
      2 million objects in the central $\sim$2 square degrees ($\sim$ 90 $\%$ of which have $i_{AB} > 24.5$)
      and becomes incomplete fainter than $i_{AB} = 25.9$ (dotted line).  The morphological catalog (the subset of the CPC 
      defining our candidate pool) contains about 195,000 objects, constrained approximately by $18 < i_{AB} < 24.5$ and exactly
      by $18 < i_{ACS} < 24.5$.  As discussed in the text, $i_{auto}$ is used throughout the paper to denote $i_{AB}$ because
      it most accurately represents the full integrated flux in the standard AB-system $i$ band.  Within 
      the magnitude range of interest ($18 < i_{AB} < 24.5$), about 80$\%$ of all CPC objects are also 
      contained in the COSMOS Morphological Catalog (CMC); the discrepancy lies largely between magnitudes 23.5 and 24.5. 
    }
    \label{niauto_f}
  \end{figure}

  AGN candidates were selected from the overlap of two catalogs: the COSMOS photometric catalog 
  (hereafter CPC) from \citet{capak07a} and the COSMOS HST Morphology Catalog (CMC) from 
  \citet{abraham04a,abraham07a}.  The former contains photometric information in \textit{uBVrizK} broadband 
  filters and photometric redshifts for 3,234,836 objects in the extended 3.5 square degree Subaru 
  optical field, and 2,326,609 objects in the central 1.7 square degrees covered by Hubble ACS imaging.  
  The central 1.7 square degrees is fully imaged with the F814W ACS filter and the resulting catalog 
  is 95$\%$ complete down to $i_{ACS} = 26.0$.  The Morphological Catalog (CMC) includes detailed 2D
  morphology for 195,706 objects restricted by $18.0 < i_{ACS} < 24.5$, which is $\sim$80$\%$ of all 
  CPC objects within the same magnitude limits.  For a plot of the differential number counts, see 
  Figure \ref{niauto_f}.  For reasons elaborated on later (see \S \ref{galcontam_ss}) we do not choose 
  to present a purely color dependent algorithm to select AGN candidate objects excluded from the CMC 
  (but part of the CPC), due mainly to heavy galaxy contamination and large increase in photometric errors. 

  In addition to the selection catalogs, we consider a ``training set'' of known AGN in the COSMOS field 
  with confirming spectroscopy \citep{trump07a}. To model the complex nature of AGN selection, we use the 
  AGN training set, along with four type 1 AGN color templates adapted from SEDs presented by \citet{budavari01a}. 
  A recent analysis of the COSMOS stellar population \citep{robin07a} is used to estimate contamination levels from 
  stars after establishing the algorithm.  A list of 1073 X-Ray point sources (hereafter XRPS) with no spectroscopy 
  are used to test algorithm efficiency after the method design has been explained.

  \subsection{COSMOS Photometric Catalog}\label{photcat_ss}

  Data are drawn from the 3 Jan 2006 data release of the Cosmic Evolution 
  Survey (COSMOS) $\sim$2 square degree equatorial field imaged with large 
  ground-based telescopes (Subaru, VLA, ESO-VLT, UKIRT, NOAO, CFHT) and 
  space-based observatories (Hubble, Spitzer, Galex, XMM, Chandra).  The 
  latest release of the photometry catalog (CPC) includes detections for 
  over 3 million objects in the Subaru $i+$ band filter in an extended 3.5 
  square degree field (the offset to SDSS $i$ is +0.3 magnitudes), and 
  magnitudes in CFH $u*, i*$ (hereinafter denoted $u$, $i_{c}$), Subaru 
  $Bj, Vj, r+, i+, z+$ (denoted $BVriz$), Kitt Peak CTIO $Ks$, narrow-band 
  Subaru $NB816$, and $F814W$ HST ACS band ($i_{ACS}$).  The CPC's main use 
  by the COSMOS collaboration has been to survey the galaxy population, of 
  which over 2 million galaxies have been detected out to $z \sim 3$ 
  \citep{scoville07a,capak07a}.  Imaging in F814W with Hubble ACS provides 
  sufficiently deep data for reliable morphological classification down to 
  $i_{ACS} > 24.5$, described more fully in \S \ref{morphcat_ss}.
  
  Photometric redshifts are estimated via two methods$-$the COSMOS team code 
  \citep{mobasher06a}, and the Baysian Photometric Redshift (BPZ) code 
  \citep{benitez99a}.  The dispersion in photometric redshifts is comparable 
  and small in either case ($\sigma(z)/z \sim 0.04$), but the Mobasher code 
  measures reddening and does a better job of breaking redshift degeneracies.  
  Although the photometric redshifts are effective for Hubble-typing galaxies, 
  they are clearly inappropriate for our AGN candidates, which have complex, 
  multi-component spectra not easily characterized by SED fitting based on 
  stellar populations.  Neither photometric redshift code uses AGN templates.  
  We will use the photometric redshifts to quantify galaxy color properties 
  and understand the contamination rates as a function of redshift, for which 
  both methods (Mobasher and BPZ) are reliable and produce similar results.
  
  The photometric catalog quotes a detection band magnitude, $i_{auto}$, 
  which defaults to Subaru magnitudes in $i+$ except in the case where the 
  source is saturated or missing in the Subaru image and CFHT $i*$ is used 
  instead. The subscript $auto$ refers to the SExtractor AUTO aperture used 
  to calculate magnitudes inside an adjustable, elliptical isophote 
  \citep{bertin96a}. CFHT magnitudes dominate $i_{auto} < 20.1$ (these 
  are saturated sources in Subaru photometry), and constitute a smaller 
  population of objects at fainter magnitudes, out to 24.  For objects with 
  photometry in both bands, the CFHT and Subaru magnitudes are consistent 
  out to $i = 26.0$, fainter than the AGN candidates which are limited by 
  the depth of the morphological catalog ($i_{auto} \sim 24.5$).  With this 
  understanding, we will not distinguish between them and will operate in 
  terms of apparent magnitude $i_{AB}$, which throughout this paper will 
  refer to $i_{auto}$.  All other magnitudes in the catalog are calculated 
  using a SExtractor fixed aperture with $3\arcsec$ diameter and are only 
  used when discussing colors.
    
\input{tab1}

  Since optical AGN have an intrinsic spread in their spectral energy 
  distributions (SEDs), the difficulty in selecting candidates is 
  aggravated by photometric errors.  For each band that we use during 
  object selection, we quote two characteristic limiting magnitudes: the 
  first is the magnitude at which the error is 10$\%$, and the second 
  is the 95$\%$ limit of catalog completeness.  Table \ref{table1} shows 
  these magnitudes for each filter.  Although the deepest band in the 
  catalog is $B$, and the other bands progressively become shallower at 
  redder wavelengths, $i+$ is chosen as the detection band image because 
  it does not bias against higher redshift objects (except at z$>$5), is 
  not effected sustantially by reddening, is the deepest red band, and is 
  typically used as the detection band for large optical surveys.  A 
  $\chi^{2}$ band \citep[a coaddition of $i$ band, $r$ band and $B$ band; 
  see][]{capak07a} has increased sensitivity and pan-chromatic advantage 
  and was also considered as a detection band; however, the $i+$ band 
  gives much better resolution needed for high quality photometric 
  calculations.

  \subsection{COSMOS Morphological Catalog}\label{morphcat_ss}
  
  The COSMOS HST morphological catalog (CMC), generated by Bob Abraham at the University of 
  Toronto \citep{abraham04a,abraham07a}, uses single filter ACS imaging to extract 2D morphology 
  classification down  to $i_{ACS} < 24.5$.  The CMC is primarily designed for use in studying 
  the morphological properties of galaxies in the COSMOS field, using 2D parametric and 
  non-parametric measures.  In the special version fo the CMC used in this project, 
  morphologies were calculated down to a level too faint for reliable galaxy work, but were 
  enabled by the fact that AGN are generally described by a point source surrounded by a 
  fainter host galaxy.  The catalog is not taken to fainter levels because the robustness 
  for even basic morphological calculations deteriorates.  The CMC includes ACS magnitudes 
  (total $AUTO$ magnitude), orientation, ellipticity, mean surface brightness, central surface 
  brightness, half light radius, signal to noise ratio, concentration index, and the Gini 
  coefficient (among other parameters not used in this study).  Since there is a significant
  color correction applied between F814W and $i_{AB}$, there is not a clean cutoff at $i = 24.5$
  for CMC sources in Figure \ref{niauto_f}.  Spurious sources from both ground based and ACS 
  data, especially in the wings of bright sources, causes the tail of few objects out to 
  magnitude $i = 26$ which should normally be included in the $18 < i < 24.5$ range.

  \begin{figure}
    \centering
    \includegraphics[width=0.99\columnwidth]{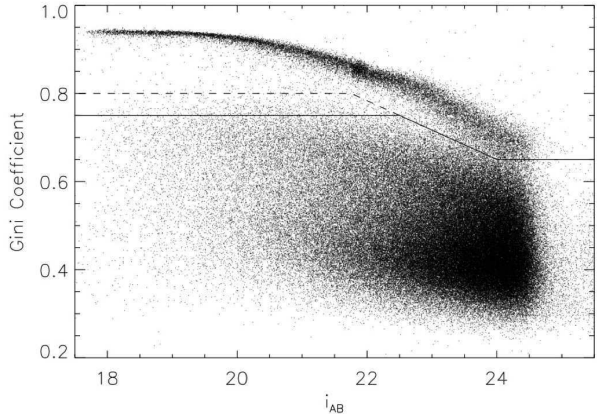}
    \caption{
      The behavior of the Gini coefficient ($G$) applied to the COSMOS/ACS i-band mosaic as a function of magnitude.
      The overall decrease in Gini with magnitude is largely due to observation bias, i.e. fainter sources are found 
      to be more extended since there is less contrast with the image background.  Altering the original Gini calculation 
      described by \citet{abraham03a} algorithm (so as $not$ to use adjustable quasi-Petrosian radii) gives cleaner 
      separation between unresolved sources (across the top), and resolved, extended sources (the bulk of the objects 
      with low $G$).  Gini is therefore useful to reject extended galaxies and retain unresolved or marginally resolved 
      stars and AGN.  The solid line indicates the $G$ selection criterion adopted later in the paper for low redshift 
      AGN (with candidates chosen to lie above the line), while the dashed line indicates the more conservative boundary 
      used for intermediate and high redshift AGN selection (see \S \ref{morphsel_s}, \S \ref{lowz_ss}, and \S \ref{highz_ss}).
    }
    \label{maggini_f}
  \end{figure}

  The Gini coefficient, hereinafter denoted $G$, is a non-parametric measure of concentration 
  ($0 < G < 1$, with $G = 1$ is a point source with all the flux in one pixel, and $G = 0$ is 
  uniformly extended with no discernable center) which doesn't assume a central pixel or a PSF.  
  The advantage of its use is that it can morphologically characterize galaxies of arbitrary 
  shape and does not require a well-defined nucleus center, which is a more general treatment 
  of PSF classification of stars and galaxies, and includes a wider scope of irregularly or 
  assymetrically shaped objects.  For a more detailed treatment and definition of Gini, as well 
  as a description of the term's origin in economics \citep{gini12}, see \citet{abraham03a}. In 
  its original context \citep{abraham04a}, Gini is calculated from pixels lying within a set of 
  quasi-Petrosian radii (unique to each object), giving the best 2D morphological analysis needed 
  for galaxy evolution studies.  At faint magnitudes ($i_{AB} > 24.5$), this approach to calculating 
  Gini breaks down, and compact objects have much lower $G$ than their bright counterparts due to 
  an inclusion of background noise within a more extended Petrosian radius.  Since this study 
  requires a clean separation of resolved and unresolved sources, we have altered the Gini 
  computation  so that the Petrosian radius is not adjusted from object to object $-$ this makes 
  the decrease in $G$ with fainter magnitude not as severe.  The behavior of $G$ with magnitude may 
  be seen in Figure \ref{maggini_f}.  The strip along the top corresponds to unresolved sources 
  (stars, compact AGN) while the large population with low $G$ corresponds to galaxies.  Clearly, 
  Gini is most useful to distinguish well-resolved galaxies from unresolved or partially resolved 
  AGN galaxies.

  \subsection{AGN Training Data}\label{training_ss}
  
  \begin{figure}
    \centering
    \includegraphics[width=0.99\columnwidth]{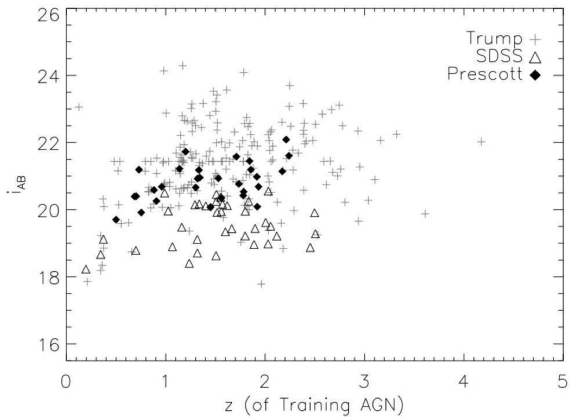}
    \caption{
      The type 1 AGN training data's magnitude as a function of spectroscopic redshift.  
      All data points are in the set of 268 type 1 AGN with color and morphology 
      information (as described in Table 2), and the different symbols represent the 
      three parent data sets: COSMOS AGN survey \citep{trump07a}, SDSS overlap with the COSMOS 
      field \citep{richards05a}, and additional spectroscopic follow up of SDSS sources on MMT 
      \citep{prescott06a}.
    }
    \label{trainingzi_f}
  \end{figure}
  
\input{tab2}

  Table \ref{table2} gives details on the AGN training data, the different sets
  of data they originate from, and the number counts of type 1 AGN, AGN included
  in the CMC and type 1 AGN included in the CMC.  Since the training data sources overlap, 
  the reader should refer to Table \ref{table2} for a breakdown of training data sources
  throughout this subsection.  Figure \ref{trainingzi_f} shows magnitude versus redshift 
  for 268 training set AGN (those training AGN for which we have both color and morphology 
  information$-$see \ref{table2}).  With 1450 total spectroscopically observed AGN targets 
  in the COSMOS field, 292 of which are type 1 AGN, we are able to infer colors and 
  morphology as a function of redshift to characterize and calibrate the AGN candidate 
  population.  The AGN training data come from four sources: an X-Ray or Radio-selection 
  from the COSMOS spectroscopic AGN survey \citep{trump07a}, the SDSS optical selection 
  with confirming spectroscopy overlaping the COSMOS field \citep{richards02a}, and
  spectroscopically confirmed SDSS optical targets from observations on MMT/Hectospec 
  \citep{prescott06a}.  The observation details on each training
  data set are given in the following paragraphs.  Since the training data sets do overlap,
  each set is described by the number of unique objects that were not included in previously
  described training data sets \citep[starting with data from][see Table \ref{table2} row 2]{trump07a}.  
  We do not use type 2 narrow line
  AGN in this paper since their optical colors have larger variations due to lower emission flux
  and obscuration.  We anticipate that the candidate objects will mostly be type 1 AGN since at 
  these magnitudes ($i \sim 24$) we need strong, broad emission features and a non-thermal 
  continuum for identification.  

  The X-Ray/Radio selected sources (limited by $i < 23$) come from the first spectroscopic 
  observations of the COSMOS AGN Survey \citep{trump07a} using the Inamori Magellan Areal 
  Camera $\&$ Spectrograph \citep[IMACS, ][]{bigelow98a} on the Magellan (Baade) Telescope.  
  The first year of observations yielded 284 AGN that were given spectroscopic redshifts,
  115 of which were originally radio sources and 169 were X-Ray sources \citep{schinnerer07a,brusa07a}.  In a second round 
  of observations, 1050 more AGN were spectroscopically confirmed in observations.
  Type 1 AGN are likely 90$\%$ complete to $i_{AB} < 23$.  The survey had 72$\%$ targeting 
  yield (the percentage of candidates that are actually AGN) down to $i_{AB} = 24$, and a 
  much better yield, $> 90$$\%$, for $i_{AB} < 22$.  A small subset of the observed targets
  was difficult to classify, but the majority was a variety of type 1 and type 2 AGN.  All 
  together, 1334 AGN were spectroscopically observed, 200 of which are type 1 AGN included 
  in the CMC (see Table \ref{table2}).  The intrinsic selection bias between 
  X-Ray/Radio selected objects and optically selected objects is valuable to investigate; 
  while working at much fainter magnitudes than the faintest SDSS optically-selected QSOs 
  ($i = 21$), we expect to incorporate the same optical selection biases.  Including the 
  X-Ray/Radio objects gives a relatively unbiased or independent sample of the optical 
  properties of the true AGN population, takes the training set to fainter magnitudes 
  ($i = 23$) than their optically selected counterparts, and may include more extended 
  well-resolved AGN galaxies which are rejected for optical selection.  

  The SDSS sample (limited by $i < 21$) comes from the overlap region of SDSS on the COSMOS 
  field (from SDSS DR1), originally targeted and selected either optically or as X-Ray 
  sources \citep[see][for selection details]{richards02a}.  Of the 86 spectroscopically 
  targeted objects (75 unique objects), 51 are type 1 broad line AGN.  These sources are primarily
  well resolved and bright ($i < 21$).  An additional 119 
  objects were optically selected as QSOs \citep{richards02a,richards04a} with high 
  confidence (90$\%$), but only 3 of these objects do not overlap with all other spectroscopic data so were not included in analysis
  \citep[including observations from][which are described in the following paragraph]{prescott06a}.

  There are 94 spectroscopically confirmed quasars (38 unique objects) in our training set observed with the
  MMT 6.5 m telescope and the Hectospec multiobject spectrograph \citep{prescott06a}.  
  The original 336 targets were marked with quasar flags drawn from the SDSS DR1 catalog, described
  by the previously discussed SDSS multicolor quasar selection algorithm.
  Eighty out of the 94 quasars did not appear in previous follow-up confirmation studies.  The quasars
  span a range of magnitudes $18.3 < g < 22.5$ and redshifts $0.2 < z < 2.3$, and the results from this study support
  the lower limit of the quasar surface density from SDSS color selection of 102 AGN per square degree down to $g = 22.5$
  over the entire COSMOS field.
  
  \subsection{Narrow Emission Line Galaxies}

  Additional observations on MMT/Hectospec from \citet{prescott06a} give 168 narrow emission line galaxies (NELGs) in the 
  COSMOS field$-$objects that were originally tagged as probable AGN from SDSS color selection but were found to be
  NELGs in spectroscopic follow-up.  Since these objects share the same colors as AGN, this set acts as a control
  for blue galaxies used to understand galaxy morphology and necessary components of the morphological selection
  design (see \S \ref{morphsel_s}).  They span redshifts $0.2 < z < 2.3$ and magnitudes $18 < i < 22.5$.  These 
  NELGs are used exclusively to understand contaminants and probe selection efficiency.

  \subsection{X-Ray Sources}\label{xrsid_ss}

  From an original set of 1865 X-Ray point sources (XRPS) in the 
  COSMOS field \citet{brusa06a,brusa07a,hasinger06a,hasinger07a,
    cappelluti07a}, the set is narrowed down to 1073 objects who 
  have 98$\%$ confidence that their optical identification is 
  secure, and are contained in the CMC \citep{brusa07a}.  They 
  are used in this paper as a test set and are treated separately 
  from the spectroscopically confirmed training set from \S 
  \ref{training_ss}.  The XRPS were not spectroscopically targeted 
  by \citet{trump07a} either because they were too faint for IMACS 
  targeting, they were not allocated a slit during observations, 
  or they lay outside regions of the 2-degree field targeted with 
  IMACS to date.  This sample is less useful in formulating the 
  algorithm designs despite its large numbers.  In terms of both 
  colors and Gini, the optical counterparts to optically faint XRPS 
  show a wide range of properties and many are low luminosity, low 
  redshift Seyfert galaxies.  Select spectroscopy reveals that 50$\%$ 
  are Type 1, 33$\%$ are Narrow Line Type 2, and 17$\%$ are ellipticals.  
  We return to this sample at the analysis stage to assess the selection 
  efficiency and completeness, and we use it roughly in the discussion 
  of the high redshift selection procedure (see \S \ref{highz_ss}).

  \subsection{AGN Templates}\label{templates_ss}
  
  The training set described in \S \ref{training_ss} gives mean AGN 
  colors out to $z_{spec} \sim 3$.  Since we intend to target AGN out 
  to $z \sim 6$, templates developed by \citet{budavari01a} are used 
  to infer colors at higher redshift.  \citeauthor{budavari01a} 
  developed four type 1 AGN templates.  Rather than characterizing 
  physical differences between AGN, these portray four optimal/empirical 
  fits to observed type 1 SEDs.  We compare the template color 
  predictions with the training data and use the best fit template 
  to predict AGN colors at higher redshift (which will be shown later 
  in Figure \ref{colorz_AGN_f}), where the training data run out. 
  Although there is significant variation in color between templates 
  (up to $\Delta$m $\sim 0.5$), there is also intrinsic spread in AGN 
  color about the mean (as seen by the training set objects, $\sigma\ 
  \sim\ 0.3$ mags), so the particular choice of template is not critical. 
  This observed color variance is a function of magnitude and thus also 
  of redshift, but we assume for simplicity that the spread about the best 
  fit template is $\sigma = \Delta m = 0.3$.  The use of the AGN color 
  templates will later be depicted graphically in Figures \ref{colorz_AGN_f} 
  (AGN color with redshift), \ref{ubz_gal_f} (galaxy colors with redshift), 
  \ref{vigini_f} (template track in $V-i$ vs. Gini), and \ref{highz_f} 
  (the 2D color selection for $z > 2.5$ candidates with template tracks 
  overplotted).

  \subsection{Stellar Surface Density}\label{stellardensity_ss}

\begin{figure}
    \centering
    \includegraphics[width=0.99\columnwidth]{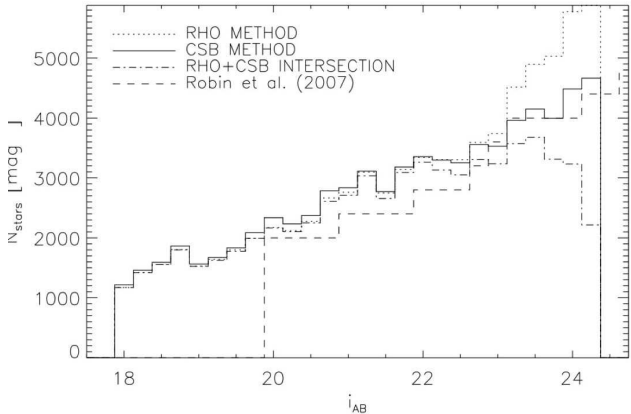}
    \caption{
      The results of stellar selection using central surface brightness (CSB) and half-light radius (RHO),
      the intersection of those two techniques, and comparing the stellar surface densities to those of the \citet{robin07a}.
      All three methodologies agree roughly with Robin's ``strict'' SED method shown here (the CSB selection working best)
      while Robin's ``loose'' SED fit method is an order of magnitude higher at the faint end, which we assert cannot be 
      representative of the true stellar surface density.  While the \citet{robin07a} method presumably isolates stars, 
      such objects are not excluded from the set of candidate AGN since their prior identification as stars cannot be certain.
    }
    \label{stellarcounts_f}
  \end{figure}

\begin{figure}
    \centering
    \includegraphics[width=0.99\columnwidth]{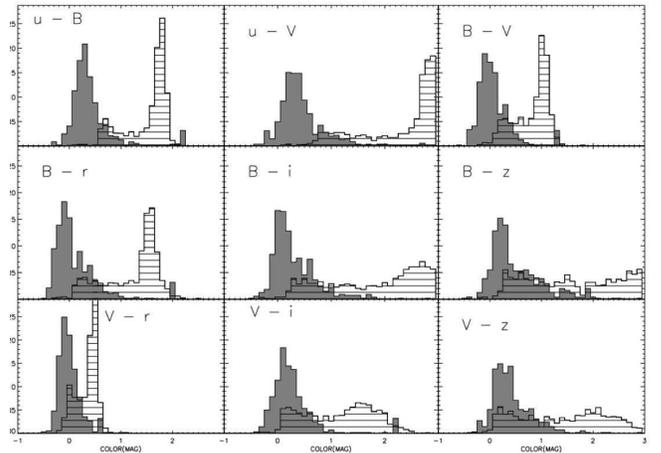}
    \caption{
      Colors of the training set of 292 AGN (solid gray) and 1791 bright stars in the COSMOS field as 
      identified by \citet{robin07a} (lined).  The distributions are normalized to equal numbers to more easily show relative
      colors.  To select $z < 2.5$ objects with the lowest stellar 
      contamination rate, a short blue baseline is used (i.e. $u-B$, $u-V$), not only because of the 
      clear separation but also because these are the deepest bands, optimal for lower luminosity AGN.
      As is discussed in the selection portion of the paper (\S \ref{algorithm_s}), the division between 
      AGN candidates and stars is taken at $u - B$ = 0.67.
    }
    \label{stlocus_f}
  \end{figure}

  A recent study of the stellar populations in the COSMOS field done by \citet{robin07a} uses HST 
  morphology coupled with detailed stellar SED fits to identify stars with 90$\%$ completeness at 
  $i=27.0$.  Their estimate (later described as the 'strict' SED fit) of the COSMOS stellar 
  population agrees well with traditional models 
  and observations of star counts \citep{chen01a,york00a,chen99a,bahcall81a,reid93a}.  This sample 
  is useful when assessing the selection algorithms' completeness and efficiency relative to 
  contaminating stars.  Their methodology identifies point sources via magnitude  and central 
  surface brightness (the $MU\_MAX$ SExtractor parameter).  We support the ``strict'' SED method 
  outlined in their paper \citep{robin07a} since the number counts procured by the SED method agree 
  with our more crude estimation of stellar counts identified solely through an identical morphological 
  point-source identification.  Robin's ``loose'' SED restriction on the quality of the SED fits 
  results in much greater stellar density counts (by a factor of ten at the faint end) and largely 
  disagree with other observations of star counts from the literature.  They include the ``loose'' 
  SED fit data to their study to demonstrate the difficulties of star/galaxy separation and gradient 
  of possible separation methods.  Our point-source identification is done in three ways: (1) magnitude 
  and central surface brightness (denoted CSB), (2) magnitude and half-light radius (denoted RHO), and 
  (3) the intersection of those two methods.  As seen in Figure \ref{stellarcounts_f}, all of these 
  methods produce roughly the same stellar surface densities as the ``strict'' stellar SED fitting method.  
  While there is the possibility that many of the AGN we are trying to target might be mislabeled as stars 
  in this set, the number counts of stars substantially outweighs the number of possibly selected AGN.  
  Since we do not use this stellar set to precisely predict colors apart from AGN (save the rough preliminary 
  estimates shown in Figure \ref{stlocus_f}) and instead use it to predict stellar number counts, the 
  inclusion of AGN is countably negligible.  In \S \ref{qlf_s}, we pass this population (identified 
  in both the CMC and CPC) through our AGN selection filters to determine the level of stellar contamination 
  as a function of magnitude, which leads to estimates of efficiency and completeness.

  \section{COLOR SELECTION}\label{colorcontam_s}

  The use of colors here differs from previous practice \citep[e.g.][]{richards02a,richards04a}
  in that efficient selection is possible without incorporating every available band into the 
  criteria.  Since quasars exhibit power law continua with a strong UV excess, the choice of
  $u-B$ to select lower redshift objects is well motivated and historically successful
  \citep{sandage65a,koo82a,warren91a,hewett95a,hall96a,croom01a,richards02a}.  Incorporating 
  the additional information from  redder baselines for low redshift, like $B-V$ and $V-r$, 
  does not improve selection efficiency on the training set, as discussed in \S \ref{lowz_ss}.
  In contrast, intermediate and high redshift selection requires a more sophisticated approach 
  since no single color is ineffective in distinguishing stars and AGN.  The optimum 2-color 
  choice for both intermediate and high redshift selection, $B-V$ and $V-i$, considers
  both the depth of the bluer bands, and the need to look towards the red
  bands for high-z candidate objects.  By selecting subsets of the catalog that represent stellar
  and galaxy populations and investigating their colors as a function of apparent magnitude,
  we conclude that the color selection method will be uniform across the entire magnitude range
  ($18.0 < i_{AB} < 24.5$).

  \subsection{AGN Colors with Redshift}\label{agncolor_ss}
  
\begin{figure}
    \centering
    \includegraphics[width=0.99\columnwidth]{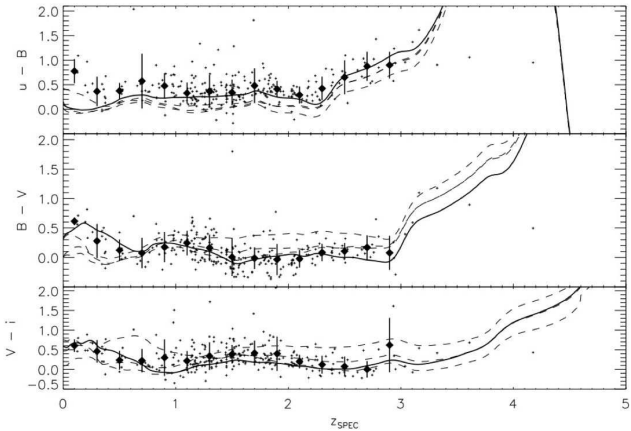}
    \caption{
      AGN colors as a function of redshift.  The 292 type 1 
      training AGN (selected as defined in the text) are small dots, the diamonds
      represent the mean of the training colors (and a 1$\sigma$ spread) taken in
      $\Delta$z = 0.2 redshift bins, and the four lines (3 dashed, 1 solid) are 
      the colors predicted from templates.  The solid line is the preferred template, as 
      judged by deviation from the mean training colors.  Although the training 
      data only extend to $z = 3$, template colors can be used to anticipate high redshift AGN colors.
      These colors rise for $z > 3$ as Lyman-$\alpha$ emission and then the Lyman limit passes from u-band to B and the redder
      bands.  A sharp drop occurs in the predicted $u-B$ color at $z \sim 4.5$ as the Lyman limit
      passes through the B band, rendering near zero flux in both bands.    }
    \label{colorz_AGN_f}
  \end{figure}

  The colors of AGN as a function of redshift are illustrated in Figure \ref{colorz_AGN_f} for 
  our three primary colors, $u-B$, $B-V$ and $V-i$.  However the spread of AGN color (from the 
  292 type 1 AGN) is consistently large in each color ($\Delta$$m \sim 0.3$).  Our low redshift 
  object selection declines rapidly in efficiency at $z \sim 2.4$ where the mean $u-B$ becomes 
  significantly redder and crosses the stellar locus as Ly$\alpha$ emission enters the B band.
  This is the natural boundary of the low redshift selection.  A similar reddening happens at 
  slightly higher redshift in $B-V$; however, this band adds selection power because AGN are 
  redder than the contaminating stars for $z > 4$.  An even redder baseline, $V-i$, shows a much 
  flatter shape as a function of redshift, and is used together with $B-V$ to optimize selection 
  of high redshift, faint candidate AGN, as described in \S \ref{highz_ss}.  
  
  The training data agree broadly with the four type 1 templates from \citet{budavari01a} up 
  to the limit of the data around $z = 3$, with the exception of the lowest redshifts in $u - B$, 
  where the AGN are redder than all templates.  Figure \ref{colorz_AGN_f} shows as a solid curve 
  the template that best fits the data, which is used for the predictions of AGN colors at 
  $z > 3$.  Colors at high redshift are inevitably uncertain.  At $z \sim 4.5$, the Lyman limit 
  passes through both u and B bands, rendering nearly zero flux in both filters and causing a 
  sharp drop in predicted $u-B$. The usefulness of each color will become more apparent as we 
  consider the contaminating populations.

  \subsection{Colors of Stellar Contaminants}\label{stellarlocus_ss}
  
  At magnitudes brighter than 21, stars are the primary contaminant in AGN selection.  
  Our goal is to classify and choose AGN candidates at all ranges of magnitudes 
  ($18.0 < i < 24.5$), so it is important to quantify stellar colors since stars
  cannot be distinguished from compact AGN morphologically, and they consist of 
  about 10$\%$ of the catalog even at the faintest levels.

  We characterize the contaminating stellar population at bright magnitudes ($i < 19$) to 
  eliminate effects from large photometric errors at faint magnitudes. This population, a 
  subset of the stars described in \S \ref{stellardensity_ss}, can be used in lieu of the 
  entire star population since we have confirmed empirically that the stellar colors do not 
  change or redden inherently as a function of apparent magnitude (for this fixed galactic 
  latitude and assuming low photometric error).  This sample has 1791 stars$-$sufficient to 
  understand color distributions.  Figure \ref{stlocus_f} shows colors of AGN and stars, 
  indicating which colors are useful in distinguishing the populations.  This reaffirms 
  $u-B$ as the best discriminator between the two populations at redshift $z \le 2.5$.  
  
  \subsection{Colors of Galaxy Contaminants}\label{galcontam_ss}
  
  Fainter than $i \sim 21$ (where our selection focuses), the overwhelming 
  majority of objects are galaxies, and therefore galaxies are the major of contaminant of the AGN 
  population.  As described by \citet{scoville07a} and \citet{mobasher06a}, all sources 
  are fit by Hubble type galaxy SEDs, yielding photometric redshifts in the range $0 <  z < 3$.  
  For the sources with the most reliable photometric redshifts (given by 
  $\chi^{2} < 25$, only the best $\sim13\%$ of the CPC), we investigate color as a function 
  of redshift for the primary contaminants: starburst and spiral galaxies.  Although 
  elliptical galaxies can theoretically be confused with high-z AGN because they are 
  compact and red, they are statistically rare and only present in the catalog in 
  significant numbers at the most recent epochs, $z < 0.4$.  Since AGN in this low 
  redshift range are much bluer than the red, compact ellipticals, we can easily reject
  ellipticals.  It is worth noting that COSMOS observations of galaxy color are available 
  for $0 < z < 3$, but photometric redshifts are not reliable at higher redshifts.
  
  \begin{figure}
    \centering
    \includegraphics[width=0.99\columnwidth]{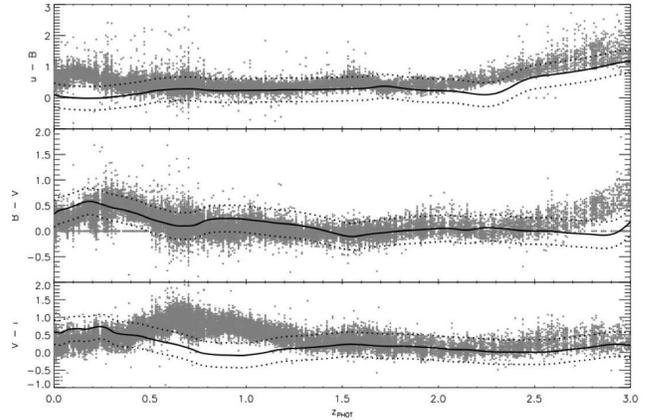}
    \caption{
      The primary color contaminant for the AGN sample are galaxies with young stellar populations: spirals and starbursts.
      Objects identified as starbursts (COSMOS catalog notation $T_{phot}$ = 5,6) with good confidence ($\chi^{2} < 25$) 
      are shown here in $(u-B)$,$(B-V)$, and $(V-i)$ colors as a function of photometric redshifts, compared to the 
      preferred AGN templates from Figure \ref{colorz_AGN_f} (solid lines with $1\sigma$ range as dotted lines).  
      Spiral galaxy colors are similar at all redshifts.  While there are small windows in redshift where the galaxy 
      colors differ from AGN, there is no way to incorporate this into the selction method since there is no prior redshift 
      indication for our candidates.
    }
    \label{ubz_gal_f}
  \end{figure}

  Figure \ref{ubz_gal_f} shows color $u-B$ as a function of photometric redshift (for 
  objects with $\chi^{2} < 25$ in the CPC) for starburst galaxies. Spirals exhibit very 
  similar colors with a slighly higher overall variance.  To clearly understand the 
  level of contamination with AGN, we have overlaid the best fit AGN template from 
  Figure \ref{colorz_AGN_f} with its 90$\%$ confidence interval illustrated by the dotted 
  lines (determined previously in \S \ref{templates_ss}).  Unlike the case for the stellar 
  population, there is little difference in color between AGN and starburst galaxies, which 
  is why we add morphological information as the basis of our low redshift selection technique.
  
  We considered the possibility of targeting very faint ($24.5 < i < 25.9$) AGN candidates (which 
  due to their faint magnitudes are not included in the CMC) using only color information, assuming 
  that the only statistically significant contaminants are faint blue galaxies.  This could work only 
  if AGN colors and galaxy colors varied over $0 < z < 3$ or if two significantly distinct colors 
  (one of them being $u-B$ to target UV excess at low redshift) showed strong separation between these 
  two populations over smaller but identical spans in redshift.  Unfortunately neither of these criteria
  is satisfied, so color selection is not effective at extremely faint magnitudes.  We also attempted 
  to target $i > 24.5$ objects using the image $FWHM$ from ground-based data, a less sensitive 
  morphological discriminator than Gini calculated using ACS data.  However, most faint objects 
  regardless of their classification by SED as stars, galaxies, or AGN have unresolved profiles with 
  $2.0\arcsec < FWHM < 2.5\arcsec$.  Since it is clear that the HST morphology is needed to target 
  faint candidates, we limit our selection to targets included in the CMC.

  \section{MORPHOLOGICAL SELECTION}\label{morphsel_s}
  
  The goal of using morphological selection is to distinguish the predominantly compact, centrally 
  concentrated AGN from the typically more extended galaxies that dominate the faint reaches of the 
  catalog.  As described in \S \ref{morphcat_ss}, the Gini coefficient is a non-parametric measure 
  of source concentration, independent of potential asymmetries or of the nature of the radial profile.  
  The Gini coefficient presents particularly strong leverage when targeting the slightly resolved AGN galaxies; 
  other methods \citep[e.g. see ][for concentration index]{abraham03a} require an assumed central pixel 
  and PSF model while Gini simply distinguishes the brightest pixels from the much fainter extended 
  component and is insensitive to the spatial arrangement of those pixels.
  
  \begin{figure}
    \centering
    \includegraphics[width=0.99\columnwidth]{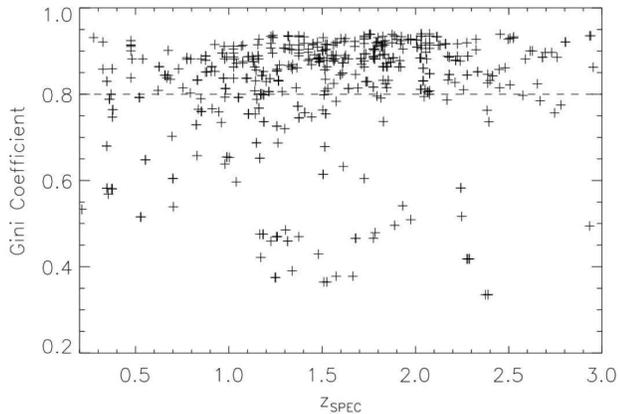}
    \caption{
      Gini Coefficient as a function of redshift for 268 type 1 AGN (with both morphological and color data)
      shows that the majority of the training set of AGN, particularly at $z > 2.0$, have $G > 0.8$, indicating essentially unresolved
      sources.  At lower redshift, a large spread in $G$ motivates a lower constraint on $G$ (also a function of magnitude as 
      shown in Figure \ref{maggini_f}).  Significant color contamination by galaxies makes it impossible to recover any 
      well-resolved AGN galaxies with $G < 0.65$.  
    }
    \label{giniz_AGN_f}
  \end{figure}

  Figure \ref{giniz_AGN_f} shows the Gini coefficient as a function of redshift
  for the 268 type 1 AGN that had confirming spectroscopy in the training set.  
  While some low and moderate redshift AGN are well resolved with $0.3 < G < 0.7$, 
  the majority ($\sim$70$\%$), particularly for $z > 1$, are unresolved with $G > 
  0.8$.  Since we have chosen to divide our selection algorithm into low, 
  intermediate and high redshifts, we will use different Gini criteria for the 
  two regimes, following the behavior illustrated in Figure \ref{giniz_AGN_f}.  
  Although it is difficult to use models to predict the behavior of Gini with 
  redshift (due to cosmic evolution and the wide range of  host galaxy properties),
  we already know that AGn are largely unresolved for $z > 1.5$.  The Gini coefficient 
  will naturally increase with redshift due to the diminishing contribution from
  the host galaxy, particularly in the $i$ band; the 4000 \AA\ break passes the $i$ band
  at $z \sim 1$.

  Figure \ref{maggini_f} showed that Gini effectively separates unresolved stars from galaxies.
  Also with the added information from Figure \ref{giniz_AGN_f} that most AGN are unresolved or 
  marginally resolved in terms of Gini, Figure \ref{maggini_f} shows the cuts made in Gini and 
  magnitude to select AGN.  At faint magnitudes, unresolved sources have lower $G$ (the right 
  end of the arc), so to include them we drop the lower limit of $G$ to 0.65 as shown.  At brighter
  magnitudes, we allow more extended or resolved sources in our low redshift candidate pool (all 
  objects above the solid line), but we set a more stringent selection for high redshift candidates 
  at higher $G$ (all objects above the dashed line) assuming high redshift AGN are less resolved 
  than their low redshift counterparts.  This distinction probably has only a small effect on 
  high-z candidate selection (since for $i < 22.5$, high-z AGN are rare).

  \begin{figure}
    \centering
    \includegraphics[width=0.99\columnwidth]{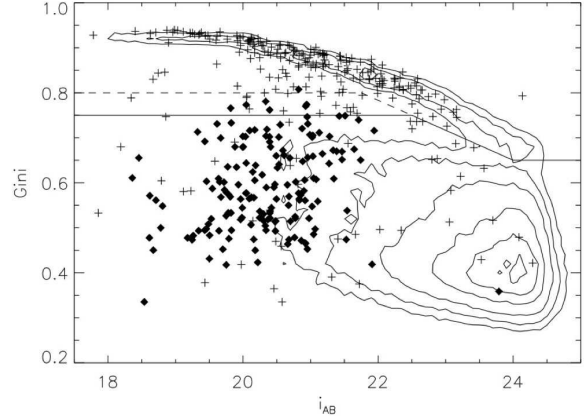}
    \caption{
      The behavior of the Gini coefficient with magnitude, as shown in Figure \ref{maggini_f}, but with contours 
      replacing the scatterplot.  The stellar locus and the region occupied by galaxies are well separated.  This 
      plot has the addition of the AGN training set (crosses) and spectroscopically confirmed Narrow Emission Line 
      Galaxies found using SDSS quasar color selection methods (diamonds).  The morphological criteria adopted for 
      this survey (shown by solid and dotted lines) exclude most NELGs while including most of the training AGN.
    }
    \label{prescottgals_f}
  \end{figure}

  The set of 168 NELGs from \citet{prescott06a} support the previously described morphological 
  selection boundaries.  The majority of the 168 NELGs (96$\%$) are well-resolved with $G < 0.75$.  
  Only 7 are accepted as AGN candidates by the low redshift Gini criteria (4$\%$) and only 1 is 
  accepted by the high-z criteria, with $G > 0.8$ ($<1\%$).  Figure \ref{prescottgals_f} indicates 
  that the majority of AGN color contaminants will be cleanly separated from AGN via morphological 
  selection.  This plot shows the same data as Figure \ref{maggini_f} (simplified to contours) but 
  also overplots the training AGN (crosses) as well as the NLEG contaminants (diamonds).  We cannot 
  use this result in a quantitative way because the statistics might be intrinsically different at
  fainter magnitudes, where NELGs may be less well-resolved and so more often confused with AGN.

  \section{AGN SELECTION}\label{algorithm_s}
  
  The primary goal of targeting the AGN population is to understand the nature of low luminosity AGN evolution 
  and spatial distribution.  The AGN selection strategy can be judged in terms of targeting efficiency and 
  completeness in recovering the predicted population.  Assessing the efficiency and completeness of our algorithm 
  depends on prior knowledge of the QLF, while also requiring additional information on the surface density of 
  contaminating stars and galaxies down to the limiting magnitude of the AGN candidates.  AGN number counts 
  have already been discussed in \S \ref{qlf_s}, and a good estimate of the stellar population is given in \S 
  \ref{stellardensity_ss}, but the galaxy contamination is the most important and unfortunately the most
  difficult to assess.  Defining the AGN selection algorithms is deeply dependent on the ability to estimate
  the efficiency of selection techniques. We used an iterative process where our selection strategy determines
  efficiency estimates (in this section, based solely on the training set) and efforts to improve efficiency 
  would alter selection methods.  In the following two subsections we describe how we define our selection 
  procedure with the motivations guiding our decisions. 

  \subsection{Low-z AGN Selection}\label{lowz_ss}

\begin{figure*}
    \centering
    \includegraphics[width=1.67\columnwidth]{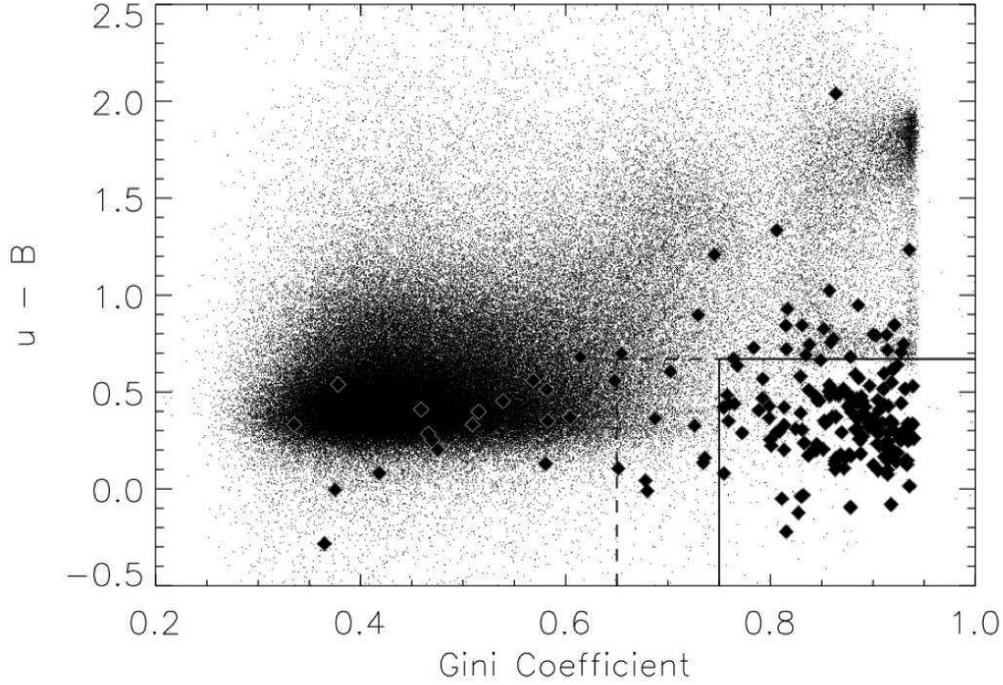}
    \caption{
      Color and morphological properties can distinguish the majority of AGN from
      their stellar and galaxy contaminants.  In this case, the separation is presented in terms of 
      $G$ vs. $u-B$.  The galaxies (thet cloud of small dots to the left)  have $G < 0.75$, the 
      stars (the smaller collection of dots at the top right) has high $G$ but primarily $u-B > 0.67$, 
      and the AGN (diamonds) are more compact than almost all galaxies and bluer than almost all
      stars.  There is a small set of well resolved AGN galaxies for which this selection is not
      effective due to heavy overlap with the blue end of the galaxy locus, primarily starburst
      galaxies, as well as a number of AGN with colors similar to the hottest stars.
    }
    \label{lowz_f}
  \end{figure*}

  The AGN population at $z < 3$ has been well studied to $i_{AB} < 22.5$, but we can target AGN 
  candidates down to $i_{AB} = 24.5$, using $G$ and $u-B$.  As previously discussed, coupling 
  these two parameters gives a clear advantage by separating galaxy, AGN, and stellar populations.  
  In Figure \ref{lowz_f}, galaxies largely have $G < 0.75$ (with starbursts at the bluer end of 
  the distribution), stars are compact ($G > 0.8$) and are relatively red with $(u-B) >\sim 0.7$ 
  and the training set AGN, represented by diamonds on Figure \ref{lowz_f}, are generally compact 
  (high $G$) and blue ($u-B < 0.7$).  Table \ref{table3} shows a flow chart of the low-z selection 
  algorithm, which is described below.

  \input{tab3}

  In this subsection (restricted to \S \ref{lowz_ss}), we define the survey's completeness and efficiency 
  in terms of the AGN training set.  Completeness is the fraction or percentage of training set AGN 
  recovered by the selection criterion, and efficiency is the number of AGN recovered or selected 
  relative to the number of star and galaxy contaminants.  Depending on our efficiency and completeness 
  goals we can vary the cuts in $G$ and $(u-B)$ to isolate AGN candidates.  The core region we use for 
  the candidate AGN pool is bounded by $0.75 < G < 1.0$ and $-0.5 < u-B < 0.6$.  It contains 51$\%$ of 
  the training data which is therefore an approximation to the selection 
  completeness for $z < 3$.  We choose 0.75 as a lower limit on Gini (rather than the more stringent choice 
  of 0.8) because there is a moderately-sized population of 20 AGN with $0.75 < G < 0.80$, and Figure 
  \ref{giniz_AGN_f} shows that the dispersion of $G$ at low redshift is high enough to warrant a lower boundary.
  The objects with marginally high $G$ could be at the Seyfert/QSO boundary with visible hosts.  Figure 
  \ref{maggini_f} highlighted the Gini selection criteria for candidates: the acceptance region for low 
  redshift objects is above the solid line.

  Another strategy would be to accept candidates with $u-B < 0.1$ and all values of $G$, to recover 
  some of the bluest and most well-resolved, low-z AGN from the training set.  Although this added 
  region does recover 7 training set AGN, there is a sharp increase in number of contaminants 
  (resulting in a 150$\%$ increase in number of candidates) as the selection skirts the blue edge of 
  the large galaxy population (primarily starbursts).  Since the anticipated gain in completeness is 
  small, only 2$\%$, we do not include this region in AGN candidate selection.

  The final aspect of the low-z selection is to set an upper bound on $u-B$.
  While AGN and stars appear to separate most cleanly for $u-B < 0.6$, many
  AGN have $0.6 < u-B < 0.9$, a region that overlaps hot main sequence stars 
  and white dwarfs.  While including this area would increase training set 
  completeness from 51$\%$ to 61$\%$, the sharp increase in overlap with 
  stars would dramatically reduce the efficiency.  The choice of $(u-B)_{max} 
  = 0.67$ was made by defining and contrasting the training set completeness 
  and efficiency.  We measure completeness for a given $(u-B)_{max}$ as the 
  fraction of training AGN within the bounds of our selection criteria out of 
  the total 292 training AGN included in the sample.  With the selection 
  criteria in terms of Gini and magnitude as in Figure \ref{maggini_f}, and 
  $u-B < (u-B)_{max}$, the completeness increases as we increase $(u-B)_{max}$.
  The efficiency (also a function of $(u-B)_{max}$) is measured as the number 
  of training AGN selected over the total number of candidates accepted by the 
  algorithm. The total number of candidates accepted includes the training set.   
  These limited definitions of completeness and efficiency are distinct from
  the expected efficiency or completeness of the overall survey, which will be 
  discussed in \S \ref{qlf_s}.

  \begin{figure}
    \centering
    \includegraphics[width=0.99\columnwidth]{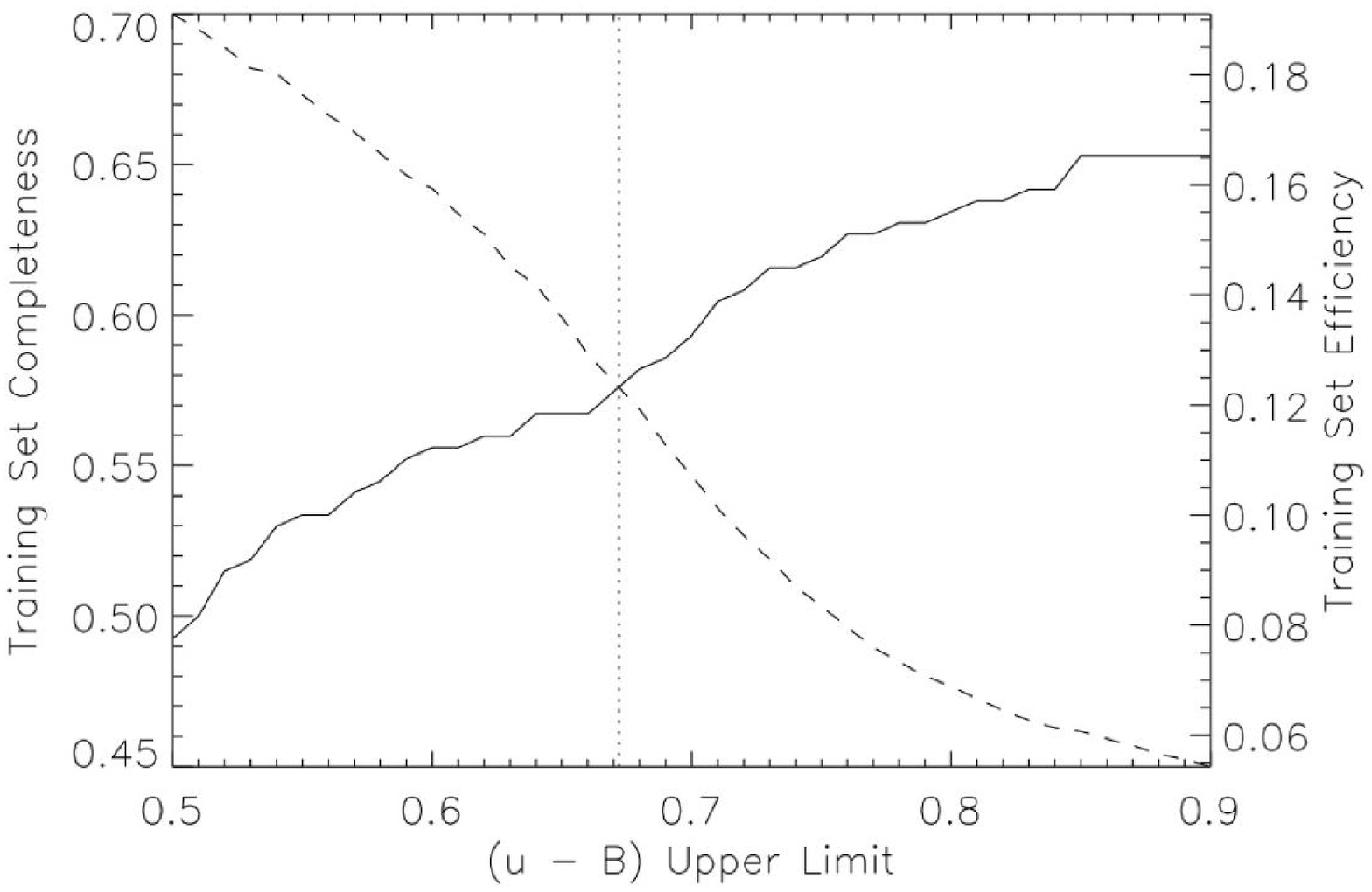}
    \caption{
      Low redshift ($z < 2.5$) AGN selection completeness from the training set (solid line) increases as a function of
      the upper bound, $(u-B)_{max}$.  The efficiency from the training set (dashed line) decreases over the same
      range of upper bounds since the region has a high density of hot stars.  The intersection of these
      two quantities occurs at $(u-B)_{max} = 0.67$ (dotted vertical line), which is chosen as the upper bound on 
      the low-z acceptance region.
    }
    \label{ub_effcomp_f}
  \end{figure}

  The region in question ($0.6 < u-B < 0.9$) contains a sizable fraction of 
  the $2.3 < z < 3.0$ AGN which are historically difficult to target. Figure 
  \ref{ub_effcomp_f} shows a plot of the training set completeness and fractional 
  training set efficiency as functions of $(u-B)_{max}$.  The training set 
  completeness increases with $(u-B)_{max}$ while the efficiency decreases.  
  Their intersection at $(u-B)_{max} = 0.67$ defines the best upper bound on 
  low-z color selection.

  Noting the evolution of AGN color as a function of redshift (Figure \ref{colorz_AGN_f}), 
  we see that beyond $z \simeq 2.5$, $u-B$ for most AGN reddens very quickly (entering the 
  stellar locus), and this selection is no longer effective.  This foreshadows why the 
  high-z selection algorithm must use more than one color.  The solid line outlines our 
  selection criteria for the brightest candidate objects ($i_{AB} > 22.5$), and the dashed 
  line extends the region to lower $G$ for fainter candidates only (the Gini-magnitude selection
  for low redshift objects is illustrated in Figure \ref{maggini_f} by the solid line).  The 
  low-z selection algorithm recovers 169 of the original 292 training AGN (58$\%$), and yields 
  a total of 2201 candidates across the 2 deg$^{2}$ COSMOS field.
    
  \subsection{Intermediate-z and High-z AGN Selection}\label{highz_ss}
  
  Unlike the low-z selection method, no single color can effectively be used to target 
  $z > 2.5$ AGN; strong contamination by the stellar locus makes that impossible.  At $z > 3$, 
  there are no training data of spectroscopically confirmed AGN, so we rely exclusively on 
  the AGN template predictions illustrated in Figure \ref{colorz_AGN_f} and number counts 
  from the X-Ray point sources with high $G$ (there are 444 above the high-z line in Figure 
  \ref{maggini_f}).  Since the X-Ray sources are not guaranteed
  to be AGN (knowing only $\sim$50$\%$ are type 1), their use in designing the selection 
  algorithm is loose, and only implimented as a guide supplimenting the use of templates.
  We group the intermediate redshift ($2.5 < z < 3.0$, dubbed int-z) selection together with the high redshift
  selection ($z > 3$) since they use the same variables and act as two subsections of a larger
  selection technique.  Below we describe this overall technique, and split into the two redshift
  regimes when it is clear that the separation is needed.

  \begin{figure}
    \centering
    \includegraphics[width=0.99\columnwidth]{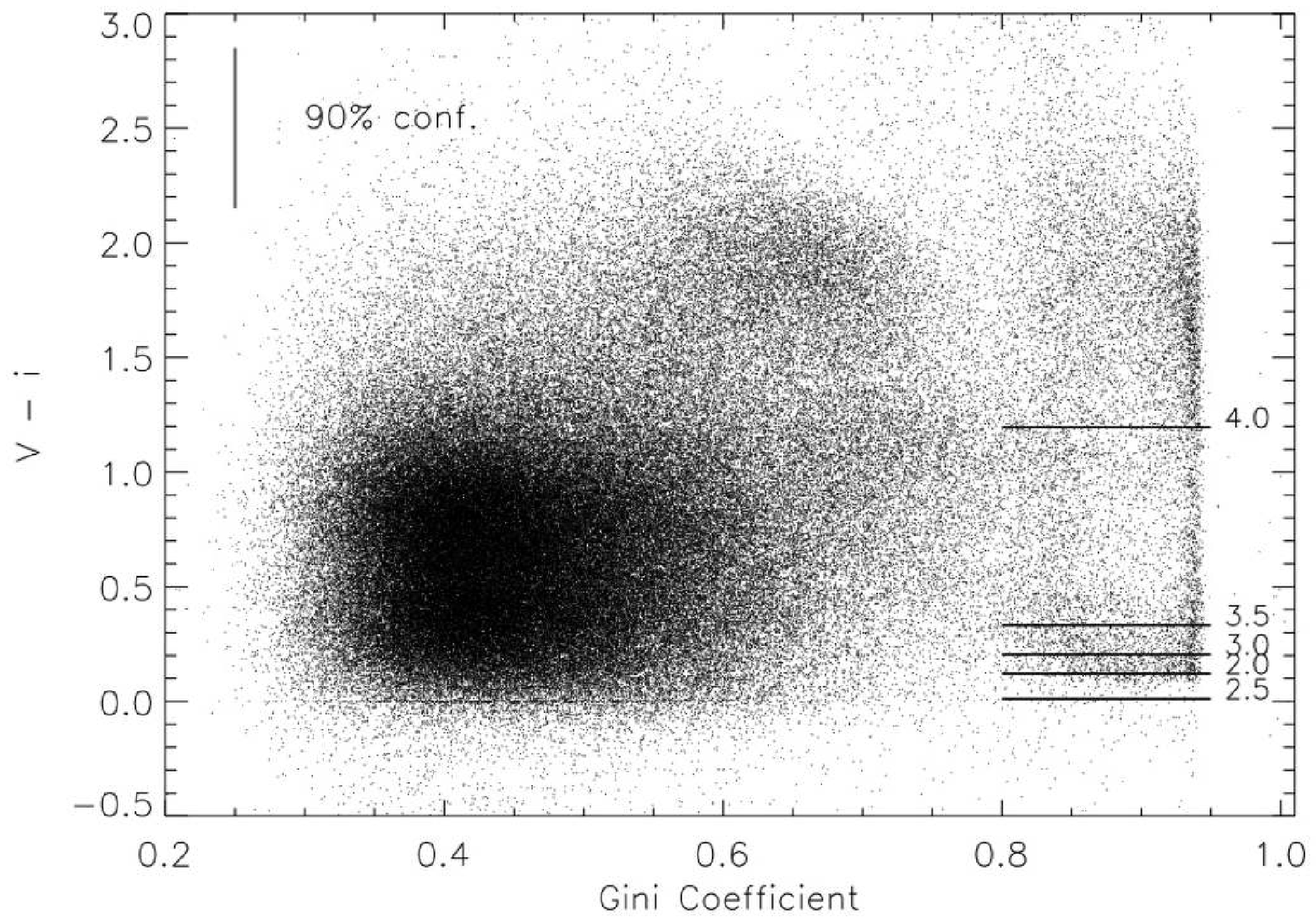}
    \caption{
      Selecting from the same catalog as for Figure \ref{lowz_f}, the use of a redder baseline and a similar selection 
      technique for high-z AGN as for low-z AGN is inefficient;  the overlap of predicted AGN colors (straight lines 
      marked with redshift) with the stellar locus is severe.  The 90$\%$ confidence interval on AGN template colors 
      is shown in the upper left. This single color technique (e.g. defining the acceptance region by $V-i < 0.1$ and 
      $G > 0.8$) rejects $z > 3.0$ AGN 60$\%$ of the time and admits large numbers of contaminating stars.
    }
    \label{vigini_f}
  \end{figure}
  
  To select objects with $z > 2.5$ we need more color information than was used to define our
  low redshift algorithm.  Figure \ref{vigini_f} shows the expected $V-i$ color of high-z AGN, 
  along with the same galaxy and star populations as shown in Figure \ref{lowz_f}. Overlap with 
  the stellar locus is severe 
  in the redshift range we are targeting, and does not let up until $z \sim 5$. Although 
  we could avoid this problem by using an even redder baseline (e.g. $V-z$, $r-z$, or 
  $z-K$), the limited depth of the catalogs and the photometric errors in these bands 
  make faint, high redshift AGN selection impossible.  Instead, we incorporate a second 
  optical color, $B-V$, which goes much deeper than the redder bands (refer to Table 
  \ref{table1} for limiting magnitudes) and which, when coupled with $V-i$, shows 
  promising separation between the stellar locus and AGN template color predictions for 
  $z > 2.5$.  To step through the stages in the int-z and high-z selection the reader should refer 
  to Table \ref{table4}.

  \input{tab4}

  \begin{figure*}
    \centering
    \includegraphics[width=1.33\columnwidth]{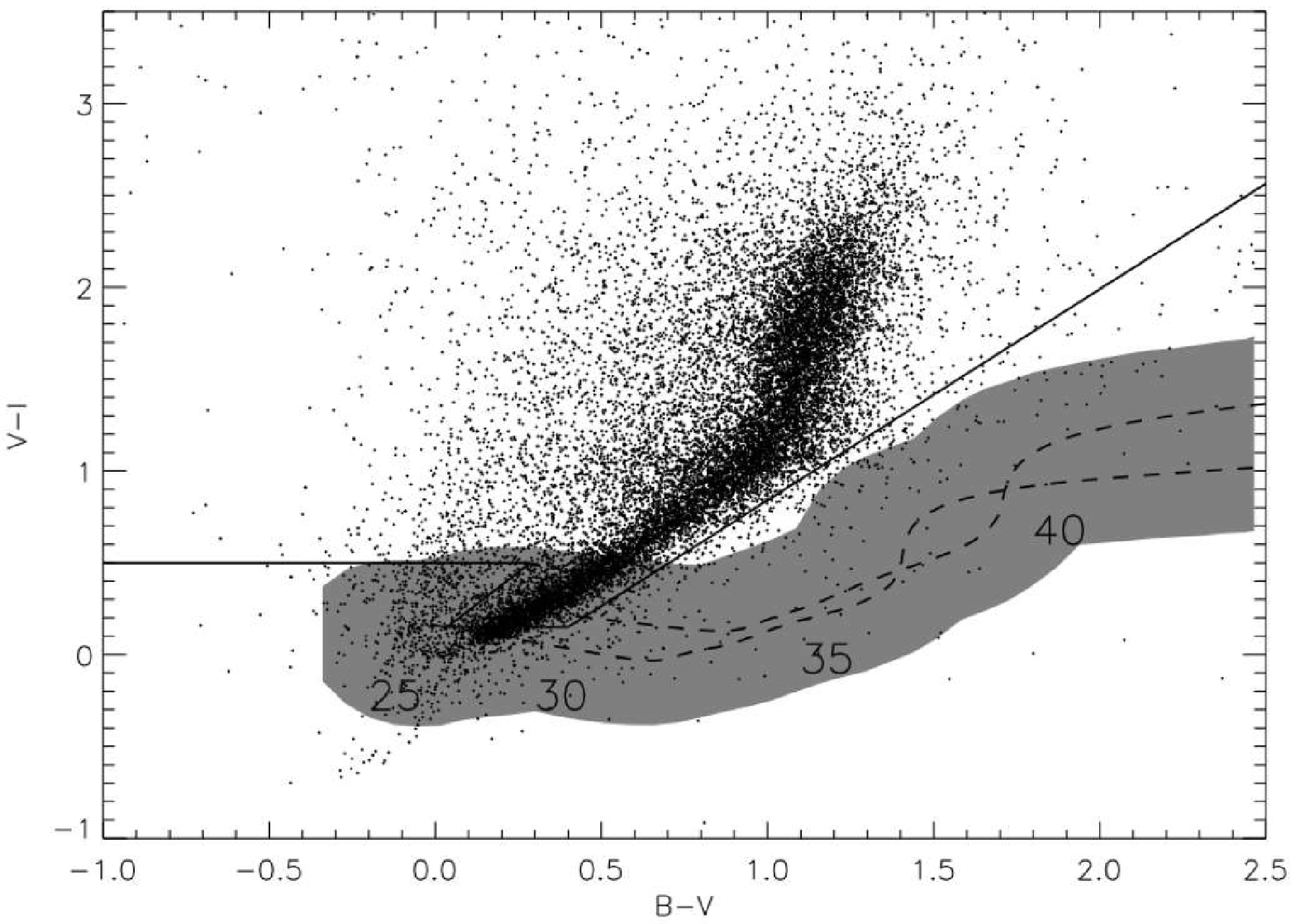}
    \includegraphics[width=1.35\columnwidth]{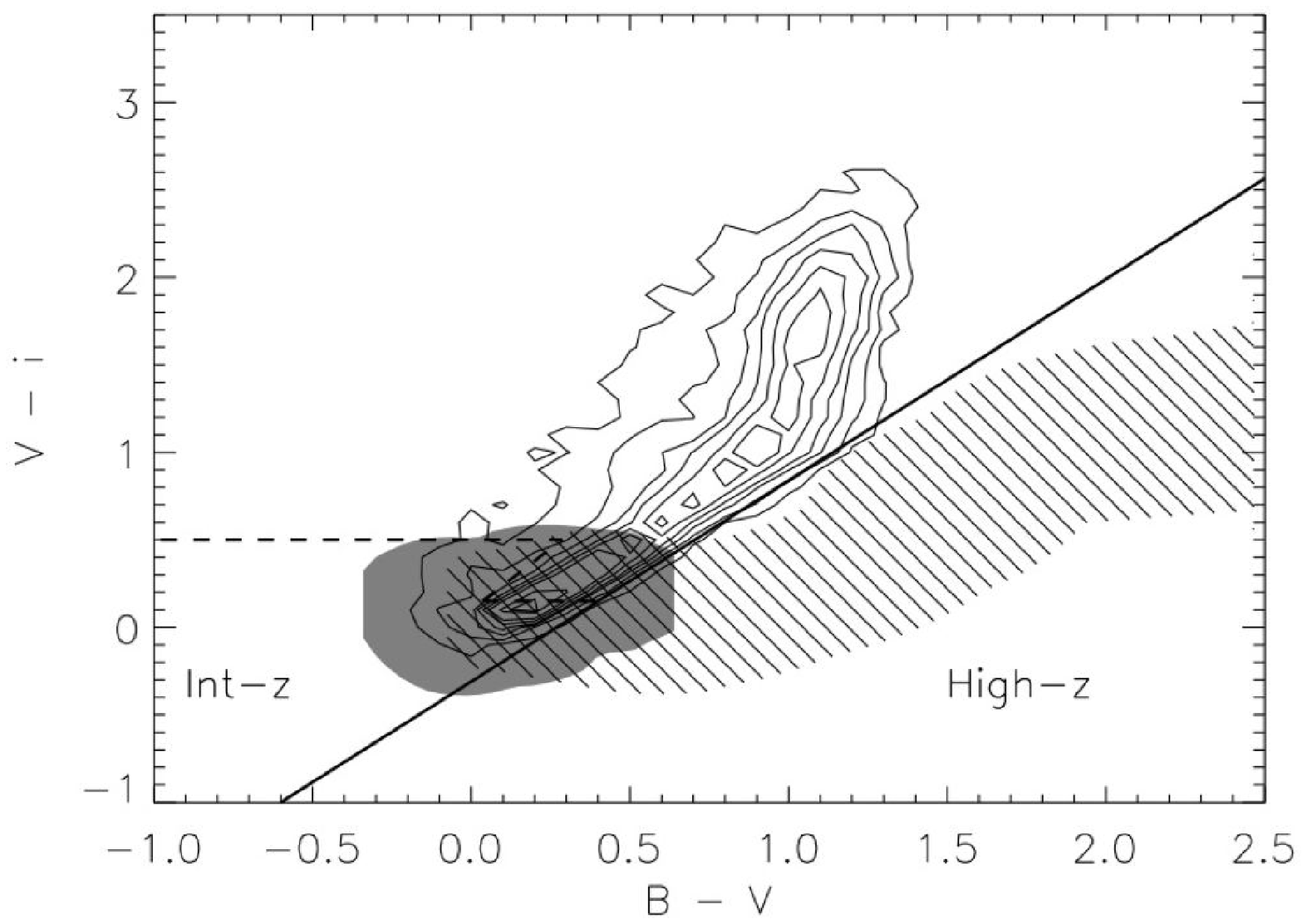}
    \caption{
      $B-V$ and $V-i$ are used to select AGN with $z > 2.5$, with the addition of a prior cut on Gini 
      shown in Figure \ref{maggini_f} by the dashed line.  Without any training data above $z = 3.0$, 
      templates are used to extrapolate the colors of AGN at high redshift.  The two best-fitting type 
      1 AGN templates (relative to low z AGN in the same colors) are shown in the top panel (dashed lines) and marked 
      at redshifts 2.5, 3.0, 3.5 and 4.0 by ``25'', ``30'', ``35'', and ``40.''  The striped region is the 
      90$\%$ envelope for potential AGN colors, adopted from the variance in $B-V$ and $V-i$ of the training 
      set AGN (for lack of better information at high redshift, variance is assumed to be constant).  
      The bottom panel shows contours converted
      from the scatter plot in the upper panel, and the template regions divided into intermediate redshift
      (gray shaded) and high redshift (striped) regions.  The solid line is the upper bound on the acceptance
      region of the high redshift selection, while the dashed line represents the upper bound of the 
      intermediate redshift selection (bound on bottom by the solid line).  X-Ray Point sources are used 
      roughly to guide selection, but on these plots their distribution would resemble a scatter plot adding 
      little visual information.  
    }
    \label{highz_f}
  \end{figure*}

  The goal of the selection algorithm is to define the optimal AGN color domain without accepting 
  significant numbers of stellar contaminants.  
  At high redshift, AGN are generally redder, fainter, and more likely to be unresolved.  
  This final assumption is based on both our small training set behavior, and the physical 
  and observational constraints at high redshift.  Therefore (coupled with data shown in 
  Figure \ref{giniz_AGN_f}), we require $G > 0.8$ for intermediate and high redshift candidates. The full 
  high-z $G$ acceptance area is shown in Figure \ref{maggini_f} as the region above both 
  the solid and dashed lines; these objects make up the initial int-z and high-z candidate sample, 
  and they are shown in the upper panel of the ($B-V$)($V-i$) diagrams in Figure \ref{highz_f}.  The gray area 
  represents a 90$\%$ envelope of all AGN template predictions for $z > 2.5$ (the redshifts 
  are marked  by ``25'', etc.).  The central lines indicate the two best fit AGN template 
  paths through the color plane (best fit lines to $B-V$ and $V-i$ as shown in Figure 
  \ref{colorz_AGN_f}).  The bottom panel of Figure \ref{highz_f} shows the int-z and high-z
  selection methods in relation to template predictions and the stellar locus, and are described
  sequencially below.   The area shaded by horizontal lines in the bottom panel of Figure 
  \ref{highz_f} represents the $z > 3.0$ AGN population outlined by templates (our high-z selection), while the area
  shaded by diagonal lines (spaced widely) represents the $2.5 < z < 3.0$ AGN population (our int-z
  selection).  The stellar locus is converted into a contour plot to schematically show areas of 
  high stellar contamination.  Compact X-Ray point sources, while not guaranteed to be AGN, are 
  overplotted as diamonds to supplement the areas highlighted by templates.

  For efficient selection at $z > 3$ we make a diagonal cut in the two color diagram, described 
  by the line $V-i < 1.15\times (B-V) - 0.31$ (shown in the plot as a heavy solid line).  The 
  region below this line contains 594 objects and 54 X-Ray sources (12$\%$ of all 444 X-Ray 
  sources considered for high-z selection), and a significant portion of the 90$\%$ AGN color 
  envelope for $z > 3$, justifying the criterion for the high-z candidate pool.  
  There are 594 high redshift objects in this selection area, 13$\%$ of which are X-Ray sources.
  After removing all X-Ray sources and training data from the selected objects, there are 515 candidates
  for the high-z selection algorithm.

  The intermediate redshift AGN occupy the area at the base of the stellar locus, blueward of the heavy
  diagonal line defining the high-z selection area.  Contamination in this region is increased greatly 
  by the population of hot main sequence stars and white dwarfs on the blue end of the stellar locus.  
  Since contamination is expected to be much higher at these redshifts, we treat intermediate redshift AGN candidates 
  separate from the high redshift AGN candidates we discussed in the previous paragraph.  The region 
  bounded by $(V-i) < 0.15$ and $V-i > 1.15\times (B-V) - 0.31$ (the heavy line) consists of 814 
  objects, 93 of which are X-Ray sources (21$\%$ of the X-Ray sample).  The upper limit $(V-i)_{max} = 0.15$
  was chosen in a similar way to $(u-B)_{max}$ from \S \ref{lowz_ss}, since the anticipated contamination
  rate greatly increases for redder values of $V-i$.  We include another region in the intermediate redshift
  selection by realizing that many X-Ray sources are bluer than the galaxy/star locus and that templates predict that 
  intermediate redshift AGN will occupy the $V-i < 0.5$ and $V-i > 1.4\times (B-V) + 0.1$ area.
  This adds 374 more candidates and 115 more X-Ray sources (up to 46$\%$ of the X-Ray sample).
  The total number of intermediate redshift AGN candidates is 913 after removing X-Ray sources and the
  training set (from an original 1188).  The acceptance region for int-z selection is shown on the bottom panel
  of Figure \ref{highz_f} enclosed by the heavy dashed line and solid line.
   
  One final catagory of targets is considered as potential high redshift AGN.  
  We have included 187 blue dropouts, where $i$ band is a detection, but $B$ 
  band is not.  These are interesting because of their very red color ($B-i > 
  2$) and faint magnitude ($i_{AB} \sim 24.0$), which potentially corresponds 
  to very high redshift AGN, in the range $3.4 < z < 5.5$.  There are only 638 
  blue dropouts in the entire COSMOS catalog, of which 187 satisfy the high-z 
  Gini cut described in \S \ref{morphsel_s}.  We add these objects into the 
  high-z candidate list, bringing the total number up to 702.

  \section{Estimating Population Statistics and Algorithm Efficiency}\label{qlf_s}

  Because we have not yet carried through with spectroscopic observations of our candidates, we cannot
  directly or reliably predict our algorithms' efficiency or completeness.  Instead we carefully
  construct a contextual arguement by roughly estimating quasar, star, and galaxy population statistics.
  By comparing our selection technique to other current algorithms in the literature, we
  estimate efficiency at $\sim$30-50$\%$ and completeness $>$60$\%$.  While detailed runs of Monte Carlo
  simulations could be used to estimate this completeness more precisely, that is not the focus of this paper.  Instead 
  a follow-up paper detailing the yeilds of this study will more carefully explore our method's robustness 
  in choosing low-luminosity AGN or potentially faint high redshift AGN in the future.

  \begin{figure}
    \centering
    \includegraphics[width=0.99\columnwidth]{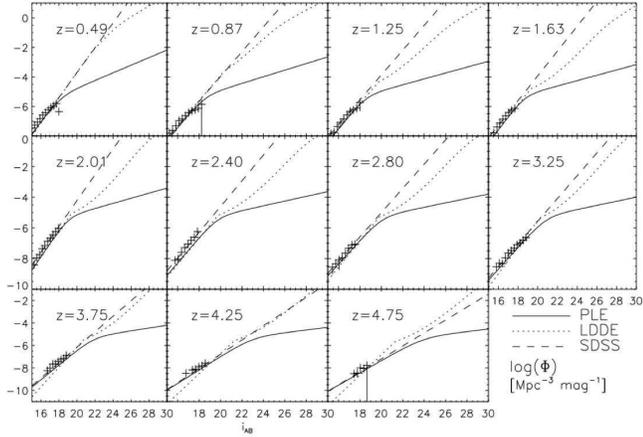}
    \caption{
      The Quasar Luminosity Function behavior at selected redshifts for the PLE model (solid), LDDE model (dotted),
      and the bright end QLF derived by \citet{richards06a} (dashed) from the SDSS DR3 quasar counts (crosses).  
      These were calibrated through bolometric corrections to agree at the bright end (the realm of SDSS data) 
      with divergence at the faint end.  The QLF models produce diverse predictions of the number counts of 
      quasars at the faint level of COSMOS.
    }
    \label{qlf_f}
  \end{figure}

\begin{figure}
    \centering
    \includegraphics[width=0.99\columnwidth]{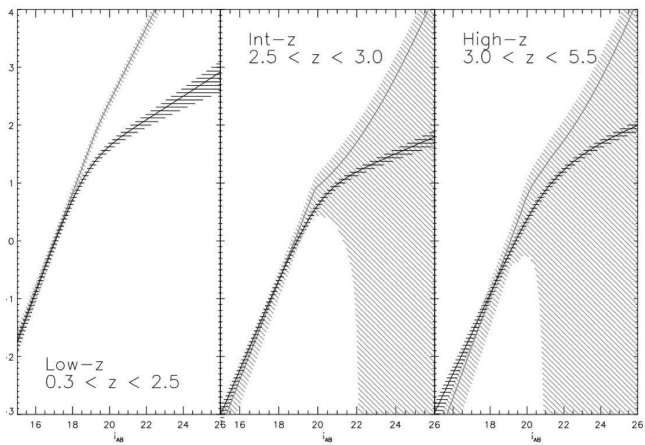}
    \caption{
      The predicted number counts of quasars in the COSMOS field (1.7 square degrees) for the low
      redshift interval ($0.3 < z < 2.4$), the intermediate redshift interval ($2.5 < z < 3.0$),
      and the high redshift interval ($3.0 < z < 5.5$).  
      The PLE quasar number counts are given by the black line, with the 1$\sigma$ margin of 
      error as the horizontally shaded regions.  The LDDE model is represented by the gray line
      with the 1$\sigma$ diagonal shaded region; at high redshifts and at faint magnitudes, the LDDE
      quasar number count is not well constrained and does not have a lower limit.  This reinforces 
      the need to target faint AGN$-$so the QLF may be more well constrained in regimes where little 
      data exists today.
    }
    \label{nqlf_f}
  \end{figure}

  The Quasar Luminosity Function (QLF) has been studied extensively over the past decade 
  with an increasing range of statistics used to verify some descriptive functional forms, 
  like the double power law \citep[e.g.][]{pei95a,peterson97a,boyle00a,croom04a}, which may be used
  to generate a prediction of the faint end QLF out to $z \sim 6$.  The most useful treatments 
  and observational insights into the QLF in the literature turn out to be inappropriate for the 
  range of magnitudes we need \citep{richards05a,richards06a,jiang06a}.  We use the pure 
  luminosity evolution (PLE) model of \citet*{hopkins07a}, who tied together several data 
  sets for the best faint end reliability, where most of our tarets lie.  The luminosity-dependent
  density evolution (LDDE) model is also often used to describe the QLF, however, at the faintest
  magnitudes we suspect it overestimates the quasar counts by $\sim$2dex and is inappropriate in this 
  context.  The predicted QLFs (PLE and LDDE)
  are shown in Figure \ref{qlf_f}\footnote{See equations 8-10, 17-20 and Table 3 of \citet{hopkins07a} for 
  the details of the PLE treatment, and equations 11-16 and Table 4 for LDDE.  
We chose the ``FULL'' model (as it is called therein) because it is best 
  across all magnitude ranges, and takes all quasar count data from all magnitudes into consideration.}.
  The predicted AGN number counts for the COSMOS field (from these QLFs at various redshifts) are shown in Figure 
  \ref{nqlf_f}, with 1$\sigma$ errors propagated from the error in the QLFs. 

  \begin{figure}
    \centering
    \includegraphics[width=0.99\columnwidth]{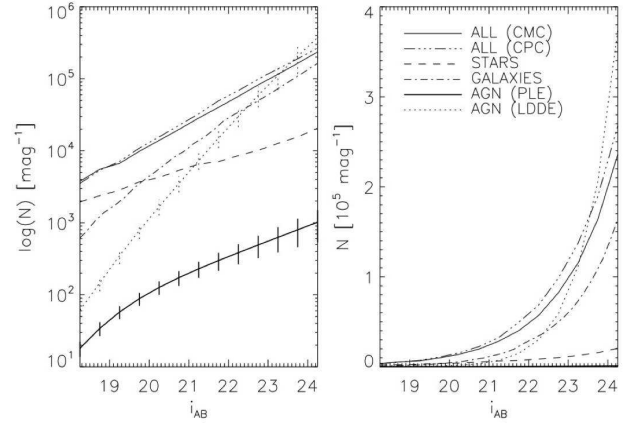}
    \caption{
      Logarithmic and linear plots of the expected number counts of objects 
      in the candidate pool.  The solid line represents all objects in the 
      CMC, while the triple-dot-dashed line shows the parent distribution of 
      objects in the CPC of the same magnitude range (see Figure \ref{niauto_f}).  
      The stellar population identified by \citet{robin07a} is shown as the 
      dashed line and at the faintest magnitudes, constituting about 10$\%$ of 
      the CMC contents.  An estimate of galaxy counts is shown as the dot-dashed 
      line and is given in \citet{capak07a} referencing previous galaxy count 
      work from COSMOS, H-HDF-N, HDF-N, HDF-S, Herschel, SDSS, CFDF, and CFHT 
      \citep{leauthaud07a,capak04a,williams96a,metcalfe01a,yasuda01a,mccracken03a,
        mccracken07b}. The predicted QLF counts (from Figure \ref{nqlf_f}) are 
      shown with appropriate error bars$-$PLE as the heavy solid line and LDDE 
      as the dotted line.  The LDDE formulation is inappropriate at the faintest 
      magnitudes where it predicts that quasars would constitute the entire 
      contents of the catalog.  Therefore the PLE formulation of the QLF (of 
      order 1$\%$ of the CMC number counts) is adopted.
    }
    \label{allobj_f}
  \end{figure}

\begin{figure}
    \centering
    \includegraphics[width=0.99\columnwidth]{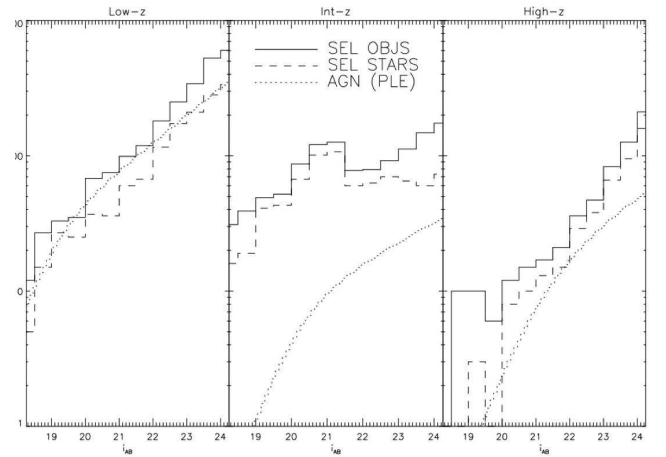}
    \caption{
      The magnitude distribution of all selected objects (solid line) through the different
      redshift algorithms.  The selected stars are shown as a dashed line, which constitute about 5$\%$ of 
      the overall stellar population from \citet{robin07a}, and $\sim$50$\%$ of all selected objects.  The QLF
      count predictions (dotted line) show the AGN numbers relative to the total number of selected objects.  
      The contamination from faint blue, compact galaxies can be inferred from this information (subtracting
      stars from the total) but there is no reliable method to measure those numbers.
    }
    \label{selobj_f}
  \end{figure}

  Figure \ref{allobj_f} gathers all predictive number counts for QSOs, stars \citep{robin07a}, and 
  galaxies \citep{capak07a} in relation to the number counts of objects in the CMC catalog.
  At the faint level of this survey, it is well known that AGN and stars are minor components of 
  a population that is made up primarily of galaxies.  The galaxy count (not previously discussed) 
  given in the latter reference is based on external measures of the galaxy surface density, which 
  agrees with several other surveys up to 80$\%$ completeness
  at $i=26.5$\footnote{e.g. the COSMOS F814W Weak Lensing Catalog \citep{leauthaud07a}, the Hawai'i Hubble Deep 
  Field \citep{capak04a}, Hubble Deep Field North \citep{williams96a,metcalfe01a}, Hubble Deep Field
  South and Herschel Deep Field \citep{metcalfe01a}, SDSS \citep{yasuda01a}, Canada France Deep Field 
  \citep{mccracken03a}, and the CFHT Legacy Survey \citep{mccracken07b}.}  This shows the overwhelming
  statistics of the galaxy population with respect to stars and AGN.  Figure \ref{selobj_f} shows the stellar
  contaminants relative to the total number of selected objects as well as predicted QLF AGN densities from 
  the PLE method.  Galaxy predictions are not shown because of their huge numbers, thus they cannot be reliably 
  determined.  Relative to the AGN counts, the stellar contaminants are roughly 3-4 times more numerous at low 
  redshift and $\sim$10 times more numerous at high redshift and hypothetically constitute $\sim$1/2 of all 
  selected candidates (although given our methodology biases against stellar selection, this is highly unlikely).

  A recent study by \citet{siana07a} presents an optical and IR selection technique of QSOs at high-z down
  to $i \sim 22$.  When coupled with results from confirming spectroscopy of $\sim$10 QSOs, they conlude a 
  completeness of 80 - 90$\%$ using detailed Monte Carlo simulations.  This estimate is based on the premise 
  that QSOs likely exhibit colors of QSO templates and are selected by their paths in color-color space, which 
  differ from stellar contaminants.  Using this methodology, we make a very similar conclusion based on the 
  similarities of our techniques: at high redshift ($z > 3$), our selection algorithm as shown in Figure 
  \ref{highz_f} will have a high completeness ($>$60$\%$).  A thorough assessment of this success rate will 
  be included in a follow-up paper detailing observational results and yields.

\begin{figure}
    \centering
    \includegraphics[width=0.99\columnwidth]{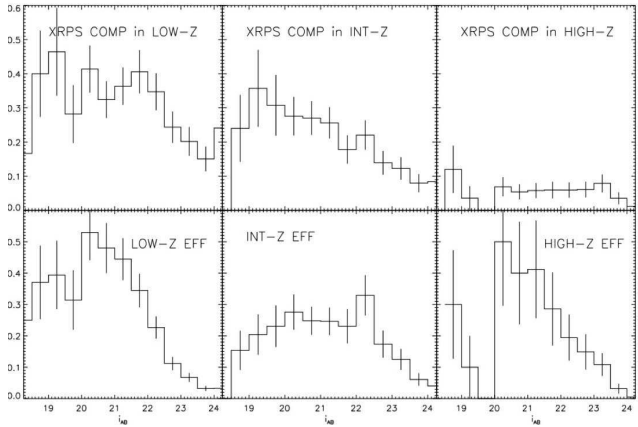}
    \caption{
      The completeness (top panels) and efficiencies (bottom panels) of the algorithms as defined by the X-Ray point sources.  
      Note that the completeness represented in each upper panel is the number of selected objects in each algorithm divided by
      1073, the total number of XRPS, and does not relate to predictions from the QLF.  While only 57 XRPS are selected in 
      high-z (low completeness for XRPS), the efficiency is quite high.  While it only is representative of a subset population
      which likely consists of low-z sources, it is valuable in understanding the true effectiveness of each technique.  
    }
    \label{xrsid_effcomp_f}
  \end{figure}

  To test the bounds on completeness and efficiency, we run the selection on the XRPS sample 
  population (which is likely comprised of 90$\%$ AGN, but only $\sim$50$\%$ type 1 AGN) and 
  compute efficiency and completeness for this sample.  The 1073 X-Ray point sources introduced 
  in \S \ref{xrsid_ss}, were not useful in designing the algorithms, but they may now be used 
  retrospectively to probe the efficiency and completeness.  We add the selected X-Ray point 
  sources back into the selected objects (modification of the last steps in Tables \ref{table3} 
  and \ref{table4}), and then compute the efficiency and completeness with respect to these X-Ray 
  objects.  Altogether, 323 X-Ray sources are targeted by the low-z algorithm and 203 are targeted 
  in int-z, and 57 using the high-z technique.  The relative completeness and efficiency of the 
  algorithms targeting the X-Ray point sources may be seen in Figure \ref{xrsid_effcomp_f}.  Low yeilds
  at faint magnitudes are potentially misleading since far fewer XRPS are at such faint magnitudes.  
  The completeness rate here must also not be misinterpreted$-$it represents the fraction of the 1073 
  XRPS which are selected by the low-z, int-z, and high-z algorithms.  Since the XRPS likely consist 
  of very few high redshift objects \citep{trump07a}, the low completeness calculation for high-z is 
  expected (upper right panel of Figure \ref{xrsid_effcomp_f}), but as shown, the algorithm is very 
  efficient in selecting those XRPS which are suspected to be high-z sources (lower right panel of 
  Figure \ref{xrsid_effcomp_f}).  The algorithms are fairly successful in selecting and targeting such 
  faint optical objects with efficiencies as high as 50$\%$ and completeness as high as 40$\%$.  The 
  X-Ray point sources are already known to be probable AGN, but this test shows that the selection 
  methodology is able to successfully target AGN with reasonable yield statistics.

  \section{Conclusions}

  The method described by this paper aims to probe the faint end of the quasar luminosity function 
  via optical AGN selection; it is framed by complex effects of a dominant contaminating population 
  of faint stars and galaxies.  Pushing optical selection to this faint level ($i \sim 24.5$) requires 
  extensive knowledge of stellar colors, stellar number counts, galaxy color contamination, compact 
  galaxies, AGN color and morphology properties, and reliable predictions of counts from the evolving 
  quasar luminosity function.  This paper establishes optical AGN selection methods for the COSMOS 
  field (with photometry from ground-based Subaru and CFHT, along with Hubble ACS imaging) using data 
  on spectroscopically confirmed AGN, X-Ray point sources, AGN color templates, and stellar studies 
  done in the COSMOS field.  We have discussed and accounted for the color of both AGN and contaminating 
  stellar and starburst galaxy populations, the use of the Gini coefficient as a reliable discriminant
  between point sources likely to be AGN or stars and extended galaxies, and the evolution of both color 
  and morphology as functions of redshift and magnitude.  While the color of blue galaxies at all 
  magnitude levels dominates the AGN contamination, leverage from the Gini coefficient can significantly
  hinder the effect of this contamination on unresolved AGN galaxies (while being defenseless to select against 
  compact blue galaxies).

  The method of targeting AGN was split into three sections: one for low redshift AGN ($z < 2.5$), one 
  for intermediate redshift AGN ($2.5 < z < 3.0$), and another for high redshift AGN ($z > 3.0$).  The 
  low and high redshift selections straddle the redshift regime of $2.5 < z < 3.0$ where AGN colors 
  resemble those of A stars in every band, and are therefore indistinguishable from stellar contaminants.  
  We design a method to target these intermediate redshift AGN, but the selection is significantly 
  hindered by increased contamination rates when compared to the low-z and high-z algorithms.  The low 
  redshift algorithm was based on the bluest baseline, $u-B$, and the Gini coefficient.  The low-z AGN 
  were identified as consistently bluer than most stars and more compact than most galaxies.  The high 
  redshift algorithm used more than one color to identify AGN, adopting $B-V$ and $V-i$.  It relied on 
  predictions from AGN templates to predict the color properties of AGN, but also used the Gini coefficient 
  to eliminate extended sources from the candidate pool. We design the intermediate redshift selection 
  as a branch of the high redshift selection; it is important to target these redshifts since it is 
  known that AGN are more numerous in the range $2.5 < z < 3.0$ than at higher redshift.   It is more
  advantageous to design a separate int-z algorithm to accept heavier contamination from stars than miss 
  this AGN population completely.
  
  The selection algorithms are designed to maximize both completeness and efficiency of selecting faint AGN.
  Although these quantities can only be estimated in advance of confirming spectroscopy of selected candidates, 
  the experiment has proven successful in its ability to recover a test sample of X-Ray point sources (which 
  are known to be $\sim$90$\%$ AGN, and $\sim$50$\%$ type 1 AGN).  With $\sim$2700 low redshift candidate 
  objects, $\sim$1000 intermediate redshift objects and $\sim$600 high redshift candidates in the 2 deg$^{2}$ 
  COSMOS field, the method could hypothetically recover $\sim$700 low-z AGN, $\sim$200 int-z AGN, and 
  $\sim$200 high-z AGN.  As a conservative estimate, roughly 2 to 10 candidates will have to be observed to 
  identify each new AGN.  A total of $\sim 160$ candidates have been observed at Magellan IMACS and LDSS3 as of
  May 2007 and more observations are planned. 

\acknowledgements
  We would like to sincerely thank the COSMOS team; information on the project is given at the public area 
  of the team website {\tt http://cosmos.astro.caltech.edu/}.  We acknowledge the staff at Caltech, CFHT, 
  CTIO, KPNO, NAOJ, STSCI, Terapix, and the University of Hawai'i for supporting this work and making the data
  available.  Additional thanks to Annie Robin of L'Observatoire de Besan{\c{c}}on, Universit{\'e} de 
  Franche-Compt{\'e} for the use of her stellar catalog, and to Andy Marble for helpful advice.  This 
  work was supported by a GO grant from STSCI.


%
%

\end{document}

%% file: tab1.tex

\begin{deluxetable}{rrr}
\tablecolumns{3} 
\tablecaption{Limiting and Completeness Magnitudes for Cosmos Photometry.} 
\tablehead{ 
\colhead{BAND}    &   \colhead{LIMITING}   &  \colhead{COMPLETENESS} 
}
\startdata 

$U_{C}$ & 26.3$\pm$0.4\phn & 27.1$\pm$0.1\phn \\
$B_{S}$ & 25.6$\pm$0.4\phn & 26.7$\pm$0.2\phn \\
$V_{S}$ & 25.6$\pm$0.5\phn & 26.6$\pm$0.2\phn \\
$R_{S}$ & 25.7$\pm$0.4\phn & 26.2$\pm$0.1\phn \\
$I_{S}$ & 24.6$\pm$0.5\phn & 25.9$\pm$0.2\phn \\
$I_{C}$ & 23.1$\pm$0.4\phn & 25.2$\pm$0.1\phn \\
$Z_{S}$ & 23.6$\pm$0.6\phn & 25.4$\pm$0.2\phn \\
$K_{K}$ & 20.1$\pm$0.3\phn & 22.9$\pm$0.1\phn \\

\enddata 
\label{table1}
\tablecomments{
The limiting magnitude is defined by the faintest magnitude (M) of which $\sigma_{M}/M < $10$\%$.
The completeness magnitude is the magnitude at which the given band is 95$\%$ complete.
The 'C' subscript corresponds to CFHT photometry, 'S' corresponds to Subaru, and 'K' corresponds to Kitt Peak CTIO.
}
\end{deluxetable}

%% file: tab2.tex
\begin{deluxetable*}{ccccccc}
\tablecolumns{6} 
\tablecaption{AGN Training Data}
\tablehead{ 
\colhead{} & \colhead{Trump} & \colhead{SDSS} & \colhead{Prescott} & \colhead{TOTAL} & \colhead{USE} \\
\colhead{(1)} & 
\colhead{(2)} & 
\colhead{(3)} & 
\colhead{(4)} & 
\colhead{(5)} & 
\colhead{(6)} \\ 
}
\startdata 
All Objects     &  1334 &  86  &  94  &  N/A  &  N/A  \\
Unique Objects  &  1334 &  75  &  38  &  1450 &  Color on all types of AGN\tablenotemark{a} \\
Type 1 AGN      &  200  &  51  &  38  &  292  &  Color on Type 1 AGN \\
AGN in CMC      &  1334 &  41  &  31  &  1406 &  Color and Morph on all types of AGN\tablenotemark{a} \\
Type 1 AGN in CMC & 200 & 37   &  31  &  268  &  Color and Morph on Type 1 AGN \\

\enddata 
\label{table2}
\tablenotetext{a}{These sets were not analyzed or used in this paper since they include type 2 AGN.}
\tablecomments{
  The AGN Training Data is broken down by source catalog \citep{trump07a,richards02a,prescott06a}, and by type
  (all AGN, type 1 AGN, AGN in CMC and type 1 AGN in CMC). Column (1) describes the type of AGN, column (2)
  represents objects from the COSMOS AGN Survey \citep{trump07a}, column (3) represents objects observed by 
  SDSS \citep{richards02a,richards05a} with confirming spectroscopy, and column (4) represents objects observed
  by \citet{prescott06a}.
  The total number of AGN of each type are given in column (5) and their use in our analysis is given
  in column (6), e.g. the most useful set has both color and morphological information for type 1 AGN
  and contains 268 objects (the redshift magnitude distribution of these objects is seen in Figure \ref{trainingzi_f}).
}
\end{deluxetable*}

%% file: tab3.tex

\begin{deluxetable*}{rrr}
\tablecolumns{3} 
\tablecaption{Low Redshift AGN Candidate Selection Process.}
\tablehead{ 
  \colhead{$N_{Obj}$}    &   \colhead{Process}   &  \colhead{Details} 
}
\startdata 

195706    & Objects in the CMC\phn & \phn \\
190316    & u, b are detections\phn & \phn \\
188553    & Brightness criterion\phn & $i_{AUTO} < 25.5$\phn \\
17697     & Gini criterion\phn & $0.65 < G < 1.0$ AND\phn \\
          & \phn & [$G > 0.75$ OR $G > -0.067\times i_{AUTO} + 2.25$]\phn \\
2370      & $u - B$ criterion\phn & $-2.00 < u - B < 0.67$\phn \\
2201      & X-Ray/Training Data\phn & Exclude XRPS and Training Sample\phn \\

\enddata 
\label{table3}
\end{deluxetable*}

%% file: tab4.tex
\begin{deluxetable*}{rrr}
  \tablecolumns{3} 
  \tablecaption{Intermediate and High Redshift AGN Candidate Selection Process
  }
  \tablehead{ 
    \colhead{$N_{Obj}$}    &   \colhead{Process}   &  \colhead{Details} 
  }
  \startdata 
  
  195706    & Objects in the CMC                 \phn &                                                     \phn \\
  192230    & Brightness Criterion               \phn & $i_{AUTO} < 25.5$                                   \phn \\
  18475     & $Gini$ Criterion                   \phn & $0.65 < G < 1.00$ AND                               \phn \\
            &                                    \phn & [$G > 0.80$ OR $G > -0.067\times i_{AUTO} + 2.25$]  \phn \\
  594       & High-z Selection in $(B-V)(V-i)$   \phn & $(V-i) \le 1.15\times (B-V) - 0.31$                 \phn \\
  515       & High-z: Remove Training Data       \phn & Exclude XRPS and Training Sample                    \phn \\
  702       & High-z: Add Blue Dropouts          \phn & Include 187 Blue Dropout Objects                    \phn \\
            &                                    \phn &                                                     \phn \\
  1188      & Int-z Selection in $(B-V)(V-i)$    \phn & $V-i > 1.15\times (B-V) - 0.31$ AND [$V-i < 0.15$ OR \phn \\
            &                                    \phn & [$V-i < 0.5$ AND $V-i > 1.4\times (B-V) + 0.1$]]    \phn \\
  913       & Int-z: Remove Training Data        \phn & Exclude XRPS and Training Sample                    \phn \\
  \enddata 
  \label{table4}
\end{deluxetable*}